\documentclass[twocolumn,showpacs,showkeys,preprintnumbers]{revtex4}

\usepackage{amssymb,amsmath,amsthm}
\usepackage{array,longtable}
\usepackage[dvips]{graphicx}
\usepackage[dvips]{color}
\newcommand{\version}{April 16, 2003}
\usepackage[mathscr]{eucal}

\renewcommand{\Im}{\mathop{\textmd{Im}}\nolimits}
\renewcommand{\Re}{\mathop{\textmd{Re}}\nolimits}

\newcommand{\rx}{\textsl{x}}
\newcommand{\Ds}{\mathscr{D}}

\newcommand{\Rh}{\mathbb R}
\newcommand{\Zh}{\mathbb Z}

\newcommand{\Ac}{\mathcal{A}}

\newcommand{\Ic}{\mathcal{I}}

\newcommand{\Hc}{\mathcal{H}}

\newcommand{\Oc}{\mathcal{O}}

\newcommand{\pd}{\partial}

\newcommand{\oz}{{\overline{z}}}
\newcommand{\ow}{{\overline{w}}}

\newcommand{\sgn}{\mathop{\mathrm{sgn}}\nolimits}

\newcommand{\app}{\alpha^{\prime}}

\allowdisplaybreaks[3]

\renewcommand{\theequation}{\arabic{section}.\arabic{equation}}

\begin{document}
\preprint{RUNHETC-2003-11}
\preprint{hep-th/0304158}

\title{Witten's Vertex Made Simple}

\author{D.M.~Belov}
\altaffiliation[On leave from]{ Steklov Mathematical
Institute, Moscow, Russia}
\email{belov@physics.rutgers.edu}
\author{C.~Lovelace}
\email{lovelace@physics.rutgers.edu}
\affiliation{
Department of Physics
\\
Rutgers University
\\
136 Frelinghuysen Rd., Piscataway, NJ 08854, USA
}

\date{\version}

\begin{abstract}
The infinite matrices in Witten's vertex are
easy to diagonalize. It just requires some
$SL(2,\Rh)$ lore plus a Watson-Sommerfeld transformation.
We calculate the eigenvalues of all Neumann matrices
for all scale dimensions $s$,
both for matter and ghosts, including fractional $s$
which we use to
regulate the difficult $s=0$ limit. We find that
$s=1$ eigenfunctions just acquire a $p$ term,
and $x$ gets replaced by the midpoint position.
\end{abstract}


\maketitle

\newpage

\tableofcontents

\section{Introduction}\label{sec:intro}
\setcounter{equation}{0}
    Witten \cite{witten} derived open strings from a field theory
with cubic Lagrangian. Its loop graphs include closed
strings \cite{freedman}. The kinetic term is the BRST operator, which
in Siegel gauge reduces to $c_0L_0$. Unfortunately the basis
which diagonalizes $L_0$ leads to forbiddingly complicated
formulae for the vertex \cite{GJ}. It was known from the beginning
that $K_1 = L_1 + L_{-1}$  commuted with Witten's vertex,
but only recently did Rastelli et al. \cite{spectroscopy} have the idea
of transforming it to the basis with $K_1$ diagonal. After a
long indirect calculation, which omitted the momenta, they found that
the Neumann matrices in the vertex take a simple diagonal
form in this basis. The present paper has two aims. Firstly
we develop a straightforward method for changing the basis.
Secondly we resolve the momentum difficulties.

    Here $L_0$, $L_{\pm 1}$ are the three generators of $SL(2,\Rh)$.
The world sheet fields in string theory belong to representations
of this group, labelled by the scale dimension $s$ (also called
the conformal weight). The string position $X^{\mu}(z)$ has $s = 0$.
If we omit the zero modes by applying $\frac{d}{dz}$, we get $s = 1$.
This was the case diagonalized by Rastelli et al. \cite{spectroscopy}.
String theory also uses $s = 2,\, -1, \frac12, \frac32, -\frac12$.
$L_0$ has discrete
  eigenvalues $m+s$, which
determine the mass spectrum
of a free string. $K_1$ has continuous eigenvalues $\kappa$, but it
is becoming clear that this basis is much more appropriate to interacting strings.

    Going from one basis to the other is essentially a
problem in $SL(2,\Rh)$ representation theory.
(However the
$s = 0$ limit is very singular.) In dealing with infinite
dimensional representations it is important to be precise
about the Hilbert space. Fortunately $SL(2,\Rh)$ has been
extensively studied by mathematicians \cite{bargmann,gelfand,ruhl}.
The appropriate Hilbert spaces for each $s$
consist of analytic functions $f(z)$ in the unit disk
and they possess a Cauchy kernel --- an integral operator $\mathrm{Id}_s(z,\oz')$ which
represents the identity and projects onto the entire space.
For any complete basis in the Hilbert
  space, this Cauchy kernel must be a sum of outer products
  of the basis functions, which fact can be used to
  normalize them. The Neumann matrices in Witten's vertex
  are the matrix elements in the $L_0$ basis of other
  integral operators \cite{peskin1}, which we call Neumann kernels.
  Now each of these kernels is just a known function of
  two complex variables. All of them can be expanded in
  outer products of $K_1$ eigenfunctions by the Watson-
  Sommerfeld contour deformation trick (familiar to
  physicists old enough to remember Regge poles). As
  expected, they are all diagonal in this basis. Dividing
  the diagonalized Neumann kernels by the Cauchy kernel
  then normalizes the eigenfunctions and immediately
  gives the Neumann eigenvalues. No matrix calculations
  are needed - one just has to check the validity of the
  contour deformation.

    This works for any scale dimension $s$, thus allowing
us to use $s$ as a regulator for the singular $s = 0$ limit.
  The exponent of the vertex contains
  a $\frac{1}{s}$ pole which enforces momentum conservation \cite{fubini}.
The situation in the $K_1$ basis is still complicated
because overlapping singularities have to be disentangled.
However we discovered a unitary transformation which separates them.
Even better, it can be reinterpreted as
  a string field redefinition
  which removes the annoying nonlocality
from Witten's
vertex, and which transforms the ghost zero mode $c_0$ into the
kinetic term conjectured for the nonperturbative vacuum \cite{rastelli}.

    Rastelli et al.\cite{spectroscopy} only considered $s = 1$, but a number
of subsequent authors \cite{marino,arevefa,dima1,Erler} extended their method
to the other cases needed in string theory. Our general
proof is an order of magnitude shorter, but we confirm
most of their conclusions. The exception is $s = 0$,
which is crucial. Our answer is simple: Including
the momentum does nothing to the continuum eigenvalues.
It adds a momentum term to their eigenfunctions,
and also a zero mode oscillator
which is just that for $L_0$ with $x$ replaced by the midpoint
position $X_L(i) + X_L(-i)$. In view of the controversy
about $s = 0$, we checked in detail that we recover the
correct Neumann matrix elements when we transform back to the $L_0$ basis.

\bigskip
    In section~\ref{sec:eigen}, we review $SL(2,\Rh)$ representations
for different scale dimensions $s$, and carefully
normalize the eigenvectors   $|m,s\rangle$   of   $L_0$   and
$|\kappa,s\rangle$   of   $K_1$. In section~\ref{sec:matter} we
diagonalize the $N$-string Neumann matrices for matter
fields and in section~\ref{sec:ghost} the $3$-string matrices for
the ghosts and superghosts, using Watson-Sommerfeld transformations.
Each
matrix is diagonalized independently, so the
identities connecting them are a cross-check.
In section~\ref{sec:zeromodes} we examine the zero modes closely.
Most of the confusion in earlier papers arose from
regularizing them inconsistently.
Section~\ref{sec:zeroNeumann} checks that
 the momentum Neumann
matrices can be correctly reconstructed.
 Section~\ref{sec:summary} is a
summary. In Appendix~\ref{app:L0}, we calculate  $L_0$   in
the new basis. Appendix~\ref{app:lemma} lists some lemmas.

\section{Normalized Eigenfunctions}
\label{sec:eigen}
\setcounter{equation}{0}
\subsection{Representations of $SL(2,\Rh)$}
It is well known that the group $SL(2,\Rh)$
is isomorphic to the group $SU(1,1)$ of quasiunitary
unimodular $2\times 2$ matrices
$\Lambda=\begin{pmatrix}
\alpha & \beta\\
\bar{\beta} & \bar{\alpha}
\end{pmatrix}$ with $|\alpha|^2-|\beta|^2=1$.
This is also known as a group of
fractional linear transformations preserving
the unit circle:
\begin{equation}
T_{\Lambda}(z)=\frac{\alpha z+\beta}{\bar{\beta} z+\bar{\alpha}}.
\label{z'z}
\end{equation}
Its unitary
representations were first considered by Bargmann \cite{bargmann}.
Good references are \cite{gelfand} and \cite{ruhl}.
The important representations for string field theory are
the discrete series $\Ds^+_s$, because they can be realized
on analytic functions
(Here $s$ is the scale dimension). The representations $\Ds^+_s$
are single valued only for $s=1,2,\dots$, while for half integer $s$
they represent the double covering group.
Unitarity fails at $s=0$, but we can continue
the representation using $s$ as a regulator.
An appropriate Hilbert space $\Hc_s$ consists of functions
$f(z)$ analytic inside the unit circle and square-integrable
on the boundary. The inner product is given by \cite{ruhl}
\begin{equation}
\langle g|f\rangle=\frac{1}{\pi\Gamma(2s-1)}\,\int_{|z|\leqslant 1}
d^2z\,\bigl[1-z\oz\bigr]^{2s-2}\, \overline{g(z)}f(z).
\label{norm}
\end{equation}
The apparent singularity at $s=\frac12$ is spurious \cite{ruhl},
but there is a real one at $s=0$. Note the two-dimensional integral.
We need a well-behaved adjoint, so we do not use analytic conformal
field theory.

The representation of the group \eqref{z'z} on this Hilbert
space is of the form
\begin{equation}
(T_{\Lambda}f)(z)=(\bar{\beta} z+\bar{\alpha})^{-2s}\,
f\left(\frac{\alpha z+\beta}{\bar{\beta} z+\bar{\alpha}}\right).
\label{trans}
\end{equation}
In other words $f(z)$ is a form of degree $s$: $f(z)(dz)^s$
is invariant under the transformation \eqref{z'z}.
One can easily check that the inner product \eqref{norm}
is invariant with respect to transformation \eqref{trans}, i.e.
$\langle T_{\Lambda}g|T_{\Lambda}f\rangle=\langle g|f\rangle$. Hence
$T_{\Lambda}$ represents a unitary operator on this Hilbert space.

The Virasoro generators are \footnote{To get the usual commutation
relation one has to change $L_n\to -L_n$.}
\begin{equation}
L_n=z^{n+1}\frac{d}{dz}+(n+1)sz^n.
\label{Lnz}
\end{equation}
$SL(2,\Rh)$ is generated by $L_0,\,L_{\pm 1}$.

\subsection{Bases in $\Hc_s$}
\label{sec:2b}
\subsubsection{The discrete basis}
The usual basis diagonalizes the elliptic generator $L_0$,
which has discrete eigenvalues $(m+s)$, $m=0,1,2,\dots$. Its
eigenfunctions normalized by \eqref{norm} are
\begin{equation}
|m,s\rangle(z)= N_m^{(s)} z^m
\quad\text{with}\quad
N_m^{(s)}=\left[\frac{\Gamma(m+2s)}{\Gamma(m+1)}\right]^{1/2}.
\label{basisn}
\end{equation}
For $s=0,-\frac12,-1,\dots$, the first $1-2s$ normalization
factors are singular. We can use fractional $s$
as a regulator, getting for example
\begin{equation*}
|0,0\rangle(z)=(2 s)^{-1/2}.
\end{equation*}
This point will be pursued in Section~\ref{sec:zeromodes}.
The Casimir is $s(1-s)$, so there is a relation
between the representations with
$s$ and $1-s$, which is important for the ghosts.
In particular,
\begin{equation}
N^{(1-s)}_{m+2s-1}\,N^{(s)}_{m}=1.
\label{NN}
\end{equation}
Also
\begin{equation*}
\left(\frac{d}{dz}\right)^{2s-1}|m+2s-1,1-s\rangle(z)=|m,s\rangle(z).
\end{equation*}
Vectors on both sides of this equation are well defined for $s\geqslant \frac12$
and $m\geqslant 0$.

\subsubsection{The continuous basis}
The generator $K_1=L_1+L_{-1}$ commutes with Witten's star product
\cite{witten}, which therefore becomes simpler when it is
diagonalized \cite{spectroscopy}. It is convenient
to map
\begin{equation}
z=i\tanh w,
\label{zw}
\end{equation}
which takes the unit disk into the strip $-\pi/4\leqslant
\Im w\leqslant \pi/4$. We assume that under a map $z\mapsto w$
the vector $f(z)$ transforms
in a trivial way
\begin{equation}
f(z)\mapsto f(z(w)).
\tag{\ref{zw}${}^{\prime}$}
\label{f->z}
\end{equation}
Then
\begin{equation}
K_1=-i\frac{d}{dw}+2is\tanh w.
\label{K_1}
\end{equation}
Since this is a hyperbolic generator, its
eigenvalues are all real numbers $\kappa$. The eigenfunctions
of \eqref{K_1} are
\begin{equation}
|\kappa,s\rangle(z)=\bigl[A_s(\kappa)\bigr]^{1/2}\,
(\cosh w)^{2s}\,e^{i\kappa w},
\label{kappas}
\end{equation}
where $A_s(\kappa)$ is an important normalization constant,
which will be determined in Subsection~\ref{subsec:nrm}.

\subsubsection{Transformation matrix between
the discrete and continuous bases}
Now we relate the two bases $|m,s\rangle(z)$ and $|\kappa,s\rangle(z)$
by expanding
\begin{equation}
(\cosh w)^{2s}\,e^{i\kappa w}
=\sum_{m=0}^{\infty} V_m^{(s)}(\kappa)z^m.
\label{223}
\end{equation}
The first two terms are easily calculated from \eqref{zw}:
$V_0^{(s)}=1,\, V_1^{(s)}=\kappa$.
By \eqref{Lnz} the vectors $|\kappa,s\rangle(z)$
satisfy the equation
\begin{equation}
\left[(1+z^2)\frac{d}{dz}+2sz-\kappa\right]|\kappa,s\rangle(z)=0.
\label{eigeneq}
\end{equation}
This gives the recursion formula
\begin{equation}
V_{m}^{(s)}(\kappa)=\frac{1}{m}\left[
\kappa\, V_{m-1}^{(s)}(\kappa)-(m+2s-2)\,V_{m-2}^{(s)}(\kappa)
\right].
\label{req1}
\end{equation}
Thus the first five polynomials are
\begin{equation}
\begin{split}
V_0^{(s)}&=1,\quad
V_1^{(s)}=\kappa,\quad V_2^{(s)}=\frac{\kappa^2}{2}-s,
\\
V_3^{(s)}&=\frac{\kappa^3}{6}-\Bigl(\frac{1}{3}+s\Bigr)\kappa,
\\
V_4^{(s)}&=\frac{\kappa^4}{24}-\Bigl(\frac{1}{3}+\frac{s}{2}\Bigr)\kappa^2
+\frac{s(s+1)}{2}.
\end{split}
\label{polynom}
\end{equation}
Some useful properties
of the polynomials $V_m^{(s)}(\kappa)$
are listed in Appendix~\ref{app:lemma}.
Normalizing both bases as in \eqref{kappas} and \eqref{basisn}
gives the transformation matrix
\begin{equation}
\langle m,s|\kappa,s\rangle=
V_m^{(s)}(\kappa)\,\frac{[A_s(\kappa)]^{\frac{1}{2}}}{N_m^{(s)}},
\label{<mk>}
\end{equation}
which is unitary for $s>0$. The unitarity follows from equation \eqref{C2}.

\subsubsection{Normalization $A_s(\kappa)$}
\label{subsec:nrm}
The normalization function $A_s(\kappa)$ has to
be determined from
\begin{equation}
\langle \kappa',s|\kappa,s\rangle=\delta(\kappa'-\kappa).
\end{equation}
Transforming \eqref{norm} by \eqref{zw} we obtain with
$w=u+iv$
\begin{multline}
\langle g|f\rangle =\frac{1}{\pi\Gamma(2s-1)}\,
\int_{-\infty}^{\infty}du\int_{-\pi/4}^{\pi/4}dv\,|\cosh w|^{-4s}
\\
\times
\bigl[\cos(2v)\bigr]^{2s-2}\,\overline{g(z(w))}f(z(w)).
\label{inner}
\end{multline}
Note that the $\cosh w$ factors cancel between \eqref{kappas}
and \eqref{inner}, so conservation of $K_1$ is just translation
invariance in $u$.

For $s=1$, the normalization $A_s$ in \eqref{kappas} can
be easily obtained by direct calculation of the integral in \eqref{inner}:
\begin{equation}
A_{1}(\kappa)=\frac{\kappa}{2\sinh\frac{\pi\kappa}{2}}.
\label{A1}
\end{equation}
In other representations it is better to proceed indirectly.
By \eqref{basisn}
\begin{multline}
\mathrm{Id}_s(z,\oz')=\sum_{m=0}^{\infty}|m,s\rangle(z)\otimes\langle m,s|(\overline{z}')
\\
=\Gamma(2s)\,\bigl[1-z\overline{z}'\bigr]^{-2s},
\label{id}
\end{multline}
where $\langle m,s|(\overline{z}')
\equiv\overline{|m,s\rangle(z')}$.
Mathematically this is the Cauchy kernel which projects onto
the entire Hilbert space. Physically it is almost
the propagator for scale dimension $s$. The divergent
factor $\Gamma(2s)$ prevents it from becoming trivial for
$s\to 0$. We can calculate
$A_s(\kappa)$ by transforming \eqref{id} to the basis
\eqref{kappas}. Inserting $z=i\tanh w$ gives
\begin{multline}
\mathrm{Id}_s(z,\oz')=2^{2s}\,\Gamma(2s)
\Bigl[\cosh w\cosh \ow'\Bigr]^{2s}
\\
\times
\Bigl\{e^{w-\overline{w}'}+
e^{\overline{w}'-w}\Bigr\}^{-2s}.
\label{id2}
\end{multline}
\begin{figure}[!t]
\centering
\includegraphics[width=221pt]{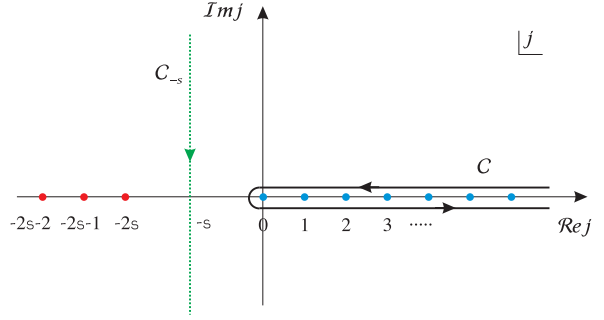}
\caption{The dots represent the poles
of the integrand in \eqref{ws1} or \eqref{M1J}. Contour
$C$ encircles the positive real
axis counterclockwise. Then we deform it to
contour $C_{-s}$, which lies parallel to the
imaginary axis at $\Re j =-s$.}
\label{fig:1}
\end{figure}
We perform a binomial expansion assuming \mbox{$\Re(\overline{w}'-w)<0$},
and then rewrite it as a contour integral
\begin{multline}
\Bigl\{e^{w-\overline{w}'}+e^{\overline{w}'-w}\Bigr\}^{-2s}
\\
=\sum_{j=0}^{\infty}
\frac{\Gamma(2s+j)}{\Gamma(2s)\Gamma(j+1)}\,
(-1)^j\,\left[e^{\overline{w}'-w}\right]^{2(s+j)}
\\
=\frac{1}{2i}\,\oint_C dj\,
\frac{1}{\sin(\pi j)}\,
\frac{\Gamma(2s+j)}{\Gamma(2s)\Gamma(j+1)}\,
\left[e^{\overline{w}'-w}\right]^{2(s+j)},
\label{ws1}
\end{multline}
where the contour $C$ encircles the positive real
axis counterclockwise (see Figure~\ref{fig:1}).
Now we use the Watson-Sommerfeld trick of
deforming the contour to lie parallel to the
imaginary axis. As $\Im j\to\pm\infty$
\begin{equation*}
\frac{1}{\sin(\pi j)}\sim e^{-\pi |\Im j|}
\quad\text{and}\quad
\frac{\Gamma(2s+j)}{\Gamma(2s)\Gamma(j+1)}\,\sim |j|^{2s-1},
\end{equation*}
so the integrand vanishes at infinity.
Since $\Gamma(z+1)$ has poles at negative integers,
the second factor in the integral \eqref{ws1} will
have poles at $j=-2s-n$ ($n=0,1,2,\dots$) and zeros
at $j=-1,-2,\dots$.
These zeros will cancel the poles of
$\bigl[\sin(\pi j)\bigr]^{-1}$ at negative integers.
So for $s>0$ we can deform the contour
to $\Re j=-s$ without  meeting any poles (see Figure~\ref{fig:1}).
However as $s\to 0$ the poles at $j=0$ and $j=-2s$
will coincide, pinching the contour between them.
Since the real axis was encircled counterclockwise, the new
integral will go down the imaginary axis.
Therefore we write
\begin{equation}
j=-s-\frac{i\kappa}{2},
\label{j1}
\end{equation}
obtaining finally
\begin{multline}
\mathrm{Id}_s(z,\oz')=
\int_{-\infty}^{\infty}d\kappa\,A_s(\kappa)\,
\bigl[\cosh w\cosh\ow'\bigr]^{2s}
\\
\times e^{i\kappa(w-\overline{w}')}
\equiv \int_{-\infty}^{\infty}d\kappa\,|\kappa,s\rangle(z)
\otimes\langle\kappa,s|(\oz')
\label{218}
\end{multline}
with
\begin{equation}
A_s(\kappa)=\frac{2^{2s-2}}{\pi}\,
\Gamma\Bigl(s+\frac{i\kappa}{2}\Bigr)
\Gamma\Bigl(s-\frac{i\kappa}{2}\Bigr).
\label{As}
\end{equation}
Using relations among $\Gamma$-functions one
can easily obtain the following recurrence formula:
\begin{equation}
A_s(\kappa)=\frac{A_{s+1}(\kappa)}{\kappa^2+4s^2}.
\label{220}
\end{equation}

By \eqref{<mk>}, $V_m^{(s)}(\kappa)$ are orthogonal
polynomials with weight $A_s(\kappa)$. The orthogonality
formula \eqref{C2} is easily checked from \eqref{C1},
which is a standard Fourier transform.

Special cases of equation \eqref{As} are
\begin{subequations}
\begin{align}
A_1(\kappa)&=\frac{\kappa}{2\sinh\frac{\pi\kappa}{2}}
\\
\intertext{which agrees with \eqref{A1} and the result of \cite{okuyama1},}
A_{\frac12}(\kappa)&=\frac{1}{2\cosh\frac{\pi\kappa}{2}}
\label{A1/2}
\\
\intertext{which agrees with \cite{arevefa}, and especially}
A_0(\kappa)&=\lim_{s\to +0}\frac{A_{s+1}(\kappa)}{\kappa^2+4s^2},
\label{A_0}
\end{align}
\end{subequations}
which requires some explanation. We did not evaluate
the limit $s\to 0$ because the expressions multiplying
$A_0(\kappa)$ may depend on $s$,
and therefore the limit can change. Some of the
following expressions will be used in our
calculations:
\begin{multline}
\lim_{s\to +0}\frac{\kappa}{\kappa^2+4s^2}=\mathscr{P}\frac{1}{\kappa},
\quad
\lim_{s\to +0}\frac{2s}{\kappa^2+4s^2}=\pi\,\delta(\kappa)
\\
\quad\text{and}\quad
\lim_{s\to +0}\frac{s^m}{\kappa^2+4s^2}=0,\quad m\geqslant 2.
\label{limits}
\end{multline}
An example of the use of these expressions is \eqref{C3'}.
In the expression for the eigenvector \eqref{kappas}
we use the square root of $A_s$, and hence we need to
know the limit $\sqrt{A_s(\kappa)}$ as $s\to 0$.
At this point it is necessary to include the sign of $\kappa$
in the definition of $\sqrt{A_0(\kappa)}$, so that
\begin{multline}
\sqrt{A_0(\kappa)}=\lim_{s\to 0}
\frac{\sgn(\kappa)}{\sqrt{\kappa^2+4s^2}}\,\sqrt{A_{s+1}(\kappa)}
\\
\equiv \mathscr{P}\,\frac{\sqrt{A_1(\kappa)}}{\kappa}.
\end{multline}

It follows from \eqref{C3}, \eqref{<mk>}
and
\eqref{220} that
\begin{equation}
\langle m+1,0|\kappa,0\rangle=\langle m,1|\kappa,1\rangle,
\label{228}
\end{equation}
so $s=1$ is just $s=0$ with $m=0$ omitted.

Equation \eqref{limits} means that for $s=0$ the spectrum of
$K_1$ includes an additional $\kappa=0$ eigenstate superimposed on
the continuum. This will be discussed
further in Section~\ref{sec:zeromodes}.
\eqref{228} also means that $s=0$ adds an $m=0$ border to $s=1$.

The ghosts and superghosts appear in the tensor product
of the representations $\Ds_s^+$ and $\Ds_{1-s}^+$. From equation \eqref{As}
one gets
\begin{subequations}
\begin{align}
\bigl(A_{n}A_{1-n}\bigr)^{1/2}&=\frac{1}{2\sinh\frac{\pi\kappa}{2}},
\label{AnAn}
\\
\bigl(A_{n+\frac{1}{2}}A_{\frac{1}{2}-n}\bigr)^{1/2}&=\frac{1}{2\cosh\frac{\pi\kappa}{2}}
\equiv A_{\frac12}(\kappa).
\end{align}
\end{subequations}

\section{Matter Vertex}
\label{sec:matter}
\setcounter{equation}{0}
\subsection{Review of the gluing vertices}
\subsubsection{Oscillator normalization}
\label{subsubsec:311}
Consider a primary conformal field $\Oc_s(z)$ of
dimension $s$. Here we assume that $2s$ is an integer,
and $\Oc_s(z)$ is a boson or fermion depending on whether
$s$ is an integer
or half integer. In the $NS$ sector the field has a mode expansion
\begin{equation}
\Oc_s(z)=\sum_{m\in\Zh-s}\frac{\Oc_m}{z^{m+s}}.
\end{equation}
We decompose it into creation and annihilation parts
with respect to the $SL(2,\Rh)$-invariant vacuum:
\begin{equation}
\Oc_s(z)=\sum_{n=0}^{\infty}N_n^{(s)}\Bigl[
\mathbf{a}_n^{+}\,z^n+
\mathbf{a}_n^{-}\, z^{-n-2s}
\Bigr]+\text{rest},
\label{32}
\end{equation}
where $N_n^{(s)}$ is defined by \eqref{basisn}, and
\begin{subequations}
\begin{equation}
N_n^{(s)}\,\mathbf{a}^{\pm}_n=\Oc_{\mp(n+s)}.
\label{oscil}
\end{equation}
The ``rest'' consists of oscillators annihilating
both the $SL(2,\Rh)$-invariant vacuum and its
conjugate (for example for $s=1$ it contains $p/z$).
We assume the following (anti)commutation relations between the oscillators
\begin{equation}
[\mathbf{a}_n^{-},\,\mathbf{a}^{+}_m]_{\pm}=\delta_{nm}.
\end{equation}
\label{bfaosc}
\end{subequations}
Although these are not the most general commutation
relations, they cover the cases in which  we are
interested: bosonic fields with $s=0$ or $s=1$
and fermionic field with $s=\frac12$.
In any case the two point correlation function
of fields $\Oc_s(z)$ on the plane  is
\begin{equation}
\langle \Oc_s(z)\Oc_s(z')\rangle=\frac{\Gamma(2s)}{(z-z')^{2s}},
\label{OOcorf}
\end{equation}
where $\Gamma(2s)$ comes from our normalization convention
\eqref{bfaosc}.

Before we proceed with formulation of a gluing vertex
let us consider a couple of examples: $s=1$ and $s=\frac12$.
Discussion of the tricky case $s=0$ we postpone to Section~\ref{sec:zeromodes}.

For $s=\frac{1}{2}$ the fermionic conformal field $\psi(z)$ has
the following mode expansion \cite{GSW}
\begin{subequations}
\begin{equation}
\psi(z)=i\left[\frac{\app}{2}\right]^{\frac12}\,
\sum_{m\in \Zh+\frac12}\frac{\psi_m}{z^{m+\frac12}}
\end{equation}
and the two point correlation function on the plane is
\begin{equation}
\langle\psi(z)\psi(z')\rangle=-\frac{\app}{2}\,\frac{1}{z-z'}.
\end{equation}
By comparing this correlation function with
\eqref{OOcorf} for $s=\frac12$ one concludes that
\begin{equation*}
i\left[\frac{\app}{2}\right]^{\frac12}\,\Oc_{\frac12}(z)=\psi(z),
\end{equation*}
and therefore
\begin{equation}
\quad\mathbf{a}_n^{\pm}=\psi_{\mp (n+\frac12)},
\quad n\geqslant 0.
\end{equation}
\end{subequations}

For $s=1$ the bosonic conformal field $\pd X(z)$ has
the following mode expansion \cite{GSW}
\begin{subequations}
\begin{equation}
i\pd X(z)=\left[\frac{\app}{2}\right]^{\frac12}\,
\sum_{m\in\Zh}\frac{\alpha_m}{z^{m+1}}
\end{equation}
and the two point correlation function on the plane is
\begin{equation}
\langle\pd X(z)\pd X(z')\rangle=-\frac{\app}{2}\,\frac{1}{(z-z')^2}.
\end{equation}
By comparing this correlation function with
\eqref{OOcorf} for $s=1$ one concludes that
\begin{equation*}
\left[\frac{\app}{2}\right]^{\frac12}\,\Oc_1(z)=i\pd X(z),
\end{equation*}
and therefore
\begin{equation}
\mathbf{a}_n^{\pm}=\frac{\alpha_{\mp (n+1)}}{\sqrt{n+1}},
\quad n\geqslant 0.
\end{equation}
\end{subequations}

\subsubsection{Continuum oscillators}
So by the unitary transformation \eqref{<mk>},
\begin{equation}
\mathbf{a}^{\pm}(\kappa)=\sqrt{A_s(\kappa)}\,
\sum_{m=0}^{\infty} \frac{V_m^{(s)}(\kappa)}{N_m^{(s)}}\,
\mathbf{a}_m^{\pm}
\label{phipm}
\end{equation}
are the oscillators in the $\kappa$-basis, satisfying
\begin{equation}
[\mathbf{a}^{-}(\kappa),\,\mathbf{a}^+(\kappa')]_{\pm}=\delta(\kappa-\kappa').
\end{equation}

Now we expand world sheet fields $\Oc_s(z)$ in these $\kappa$-oscillators.
We assume that $z$ is on the unit circle, which allows
us to change $1/z$ to $\oz$ (This is not a restriction
for us, because we are only interested
in the world-sheet fields on the boundary). Hence the expansion is
\begin{widetext}
\begin{multline}
\left.\Oc_s(z)\right|_{|z|=1}=
\int_{-\infty}^{\infty} d\kappa\,
\Bigl\{
\mathbf{a}^+(\kappa)\,|\kappa,s\rangle(z)
+\mathbf{a}^-(\kappa)\,\oz^{2s}\,|\kappa,s\rangle(\oz)
\Bigr\}+\text{rest}
\\
\equiv\int_{-\infty}^{\infty} d\kappa\,\sqrt{A_s(\kappa)}\,
\Bigl\{
\mathbf{a}^+(\kappa)\,[\cosh w]^{2s}\,e^{i\kappa w}
+\mathbf{a}^-(\kappa)\,\oz^{2s}\,
[\cosh \ow]^{2s}\,e^{-i\kappa \ow}
\Bigr\}+\text{rest}.
\label{psiz}
\end{multline}
This expansion is easy to obtain by using the
representation \eqref{218} for the Cauchy kernel.

\subsubsection{Gluing vertex}
The $N$-string vertex $\langle V^{(s)}_N|$ is a multilinear map from
the $N$-th  power of an oscillator Fock space to the complex numbers.
For a conformal field $\Oc_s(z)$ of dimension $s$ (described
in subsection~\ref{subsubsec:311})
it can be written as a Gaussian
state of the form
\begin{equation}
\langle V^{(s)}_N|={}_{1\dots N}\langle 0|\exp\left[
\frac{(-1)^{2s}}{2}\sum_{n,m=0}^{\infty}\sum_{I,J=1}^{N}\bigl(M_{s,N}^{IJ}C
\bigr)_{mn}\,
\mathbf{a}_m^{-(I)}\mathbf{a}_n^{-(J)}\right].
\label{mbra}
\end{equation}
\end{widetext}
Here
${}_{1\dots N}\langle 0|$ is a tensor product of $SL(2,\Rh)$
invariant vacua from each Fock space,
$\mathbf{a}_n^{-(I)}$ are annihilation oscillators \eqref{bfaosc}
acting in the $I$-th Fock space,
$C_{nm}=(-1)^n\delta_{nm}$ is a twist
matrix and $M_{s,N,\,\,nm}^{IJ}$ are the Neumann matrices defining the
gluing vertex \cite{GJ,peskin1}. The Neumann matrices
are symmetric or skewsymmetric
$\bigl(M_{s,N}^{IJ} C\bigr)_{mn}=(-1)^{2s}\bigl(M_{s,N}^{JI}C\bigr)_{nm}$
and
satisfy the cyclicity property $M_{s,N}^{IJ}=M_{s,N}^{I+1,J+1}$.

For any string field theory matter vertex, the Neumann matrices
can be generated from a kernel operator
\begin{equation}
\begin{split}
\bigl(M_{s,N}^{IJ}C\bigr)_{mn}&=\langle m,s|M_{s,N}^{IJ}C|n,s\rangle,
\\
\bigl(M_{s,N}^{IJ}\bigr)_{mn}&=\langle m,s|M_{s,N}^{IJ}|n,s\rangle,
\end{split}
\label{MmnIJ}
\end{equation}
where the states are \eqref{basisn}. The expression
for the operator $M_{s,N}^{IJ}C$ can by obtained using
the conformal definition \cite{peskin1} of the gluing vertex:
\begin{multline}
\bigl(M_{s,N}^{IJ}C\bigr)(z,\oz')=\Gamma(2s)\,
\\
\times\left\{
\frac{\Bigl[h_I'(z)\Bigr]^s\Bigl[h_J'(\oz')\Bigr]^s}
{\Bigl[h_I(z)-h_J(\oz')\Bigr]^{2s}}
-\frac{\delta^{IJ}}{(z-\oz')^{2s}}
\right\},
\label{CMs}
\end{multline}
where $I,J=1,\dots, N$ label the glued strings
and the maps $h_I(z)$ are defined below.
Essentially  \eqref{CMs} just generalizes
LeClair et al. \cite{peskin1} to arbitrary scale dimension $s$.
The powers of $s$ are determined by covariance under \eqref{trans}.
The denominator must match the propagator \eqref{OOcorf}.
When $s$ is fractional we assume the principal branch of
the power function, in other words the branch cut is
on the negative real axis.
The
$\Gamma$-function is needed to give a nontrivial $s\to 0$ limit.
To be consistent with our scalar product \eqref{norm}
we have put $\oz'$ as the second argument instead of $z'$.
The second term proportional to $(z-\oz')^{-2s}$ comes from the normal
ordering when one acts by two operators of weight $s$ in the same
Fock space. When using a contour integral representation
for the Neumann matrices (as in \cite{peskin1}) one never sees
this term: it simply gives zero contribution. In our calculations
this term cancels some divergences appearing in the diagonalization
of $M^{II}_{s,N}$. Gross and Jevicki \cite{GJ} (paper 3, eq. (3.28))
also mentioned that the free propagator must be subtracted from $M^{II}_{s,N}$.

Projecting with \eqref{basisn} in \eqref{MmnIJ} is equivalent to expanding
\eqref{CMs} in $z$ and $\oz'$, picking out the term $z^m\oz^{\prime\,n}$
and then dividing
its coefficient by $N_m^{(s)}N_n^{(s)}$. (This last assumes
oscillators are normalized $[\mathbf{a}_m^-,\,\mathbf{a}_n^{+}]=\delta_{mn}$.)
The contour integrals in \cite{peskin1} achieve the same result.

To obtain an expression for the operator $M_{s,N}^{IJ}$ it is enough
to notice that the twist operator $C$ acts on the eigenvectors
$|n,s\rangle(z)$ by changing the sign of the argument $z$:
\begin{equation}
\bigl(C|n,s\rangle\bigr)(z) =|n,s\rangle(-z).
\end{equation}
In other words to get an expression for $M_s^{IJ}$ one
has just to change the sign of $\oz$ in \eqref{CMs}:
\begin{multline}
M_{s,N}^{IJ}(z,\oz')=\Gamma(2s)\,
\\
\times\left\{
\frac{\Bigl[h_I'(z)\Bigr]^s\Bigl[h_J'(-\oz')\Bigr]^s}
{\Bigl[h_I(z)-h_J(-\oz')\Bigr]^{2s}}
-\frac{\delta^{IJ}}{(z+\oz')^{2s}}
\right\}
\label{Mmaps}
\end{multline}

\subsection{Diagonalizing Witten's vertices}
Different string field theories use different maps $h_I(z)$.
For Witten's $3$-string vertex
\begin{subequations}
\begin{equation}
h_2(z)=\left(\frac{1-iz}{1+iz}\right)^{2/3}=e^{4w/3},
\label{h2}
\end{equation}
where $z=i\tanh w$, and
\begin{equation}
h_{1,3}(z)=e^{\pm 2\pi i/3}\,h_2(z).
\end{equation}
\label{h3}
\end{subequations}
However we can diagonalize the $N$-string vertex for very little extra
trouble. We therefore take
\begin{equation}
h_I(z)=e^{i\varphi_I}\left(\frac{1-iz}{1+iz}\right)^{2/N}=
e^{i\varphi_I}\,e^{4w/N},
\label{h}
\end{equation}
where $z=i\tanh w$ and $\varphi_I=\frac{2\pi}{N}\,(\alpha_N-I)$.
Here $\alpha_N$ is a real number which is chosen
in such a way that all angles $\varphi_I$ lie in the
range $(-\pi,\pi]$. This last requirement is important because
we use rational powers in the definition of the Neumann matrix.
Then
\begin{equation}
h_I'(z)=-\frac{4i}{N}\,\cosh^2 w\,h_I(z).
\label{hI}
\end{equation}
As we saw in \eqref{inner} the map $z\to w$ takes the unit disk into the
strip $-\pi/4\leqslant \Im w\leqslant \pi/4$. The maps
$w\to h_I$ then transform this strip into $N$
$360^o/N$ wedges, which are glued together by the Neumann matrices
\cite{peskin1}.
Note that \eqref{Mmaps} is homogeneous in the $h_I$'s. The $\cosh w$
factors from \eqref{hI} and \eqref{kappas} always cancel against
the inner product \eqref{inner},
so homogeneity in $h$ implies translation invariance under
$w\to w+c$, $\overline{w}'\to\overline{w}'+c$, and therefore
conservation of $K_1$.

To diagonalize the $N$-string Neumann matrix, we proceed as in \eqref{id2}
et seq. --- first a binomial expansion of \eqref{Mmaps}, then a Watson-Sommerfeld
transformation. The final result is the same whichever way round
one does the binomial expansion. Thus if
$\Re (\overline{w}'-w)<0$,
\begin{widetext}
\begin{multline}
M^{IJ}_{s,N}(z,\oz')=\Bigl[
\cosh w\cosh\overline{w}'
\Bigr]^{2s}
\!\oint_C\frac{dj}{2i\sin(\pi j)}
\frac{\Gamma(2s+j)}{\Gamma(j+1)}
\\
\times\Biggl\{
\left(\frac{4}{N}\right)^{2s}
\left[-e^{i(\varphi_J-\varphi_I)}e^{4(\overline{w}'-w)/N}\right]^{s+j}
-\delta^{IJ}\,2^{2s}
\left[-e^{2(\overline{w}'-w)}\right]^{s+j}
\Biggr\},
\label{M1J}
\end{multline}
where $\varphi_J-\varphi_I=\frac{2\pi}{N}\,(I-J)$.
The contour $C$ encircles the positive real axis counterclockwise
(see Figure~\ref{fig:1}). Before deforming it as in Figure~\ref{fig:1}
we must worry about the contour at infinity.
Starting from here we will consider $I\ne J$ and $I=J$ separately.
\end{widetext}

\subsection{Matrices $M_{s,N}^{IJ}$ for $I\ne J$}

By cyclic symmetry we can fix $I=1$, then $J=2,\dots,N$.
In this case we can interpret $(-1)=e^{i\pi}$ and therefore
\begin{equation*}
-e^{i(\varphi_J-\varphi_1)}=e^{\frac{2\pi i}{N}\left(
1-J+\frac{N}{2}\right)}.
\end{equation*}
This guarantees that for $2\leqslant J \leqslant N$
\begin{equation}
-\pi <\arg\Bigl( -e^{i(\varphi_J-\varphi_1)}\Bigr)<\pi.
\end{equation}
After dividing by $\sin(\pi j)$ we
get the following asymptotic behavior of the integrand
\begin{equation*}
\sim |j|^{2s-1}\,e^{-2\pi|\Im j|/N}
\end{equation*}
as $\Im j\to\pm\infty$. Hence for $M^{1J}$ ($J\ne 1$) the
integrand vanishes at infinity.
The poles of $\bigl[\sin(\pi j)\bigr]^{-1}$
at negative integers are cancelled by zeros of $\bigl[\Gamma(j+1)\bigr]^{-1}$,
so for $s>0$ we can shift the contour to $\Re j=-s$ as in
Figure~\ref{fig:1} by writing
\begin{equation}
j=-s-\frac{iN\kappa}{4}
\label{j2}
\end{equation}
to get
\begin{multline}
M^{1J}_{s,N}(z,\oz')=\int_{-\infty}^{\infty}d\kappa\,
e^{\frac{\pi \kappa}{4}\,(2+N-2J)}
\,B_{s,N}(\kappa)\,
\\
\times
e^{i\kappa(w-\bar{w}')}\,
\bigl(
\cosh w\cosh\overline{w}'
\bigr)^{2s},
\end{multline}
where
\begin{equation}
B_{s,N}(\kappa)=\frac{1}{2\pi}\left[\frac{4}{N}\right]^{2s-1}
\Gamma\Bigl(s+\frac{iN\kappa}{4}\Bigr)\Gamma\Bigl(s-\frac{iN\kappa}{4}\Bigr).
\label{Bs}
\end{equation}
This displays the Neumann matrix as an outer product of $K_1$
eigenfunctions \eqref{kappas}. Notice also that the normalization
$A_s(\kappa)$ introduced in \eqref{As} is equal to $B_{s,2}(\kappa)$.

\subsection{Matrix $M^{II}_{s,N}$}
\label{sec:3d}

If $I=J$, the first term in \eqref{M1J} contains a factor
\begin{equation*}
(-1)^{s+j}=e^{\pm i\pi(s+j)}.
\end{equation*}
Either interpretation of $(-1)$
will cancel the nice falloff of $\bigl[\sin(\pi j)\bigr]^{-1}$ in \eqref{M1J}
and prevent deformation of the contour.
However in this case the second term in \eqref{M1J} comes into play.
The bad asymptotic behavior at \mbox{$\Im j\to\pm\infty$}
is closely related to the coincident singularity at $w=\ow'$
coming from the vanishing denominators in \eqref{Mmaps}. Thus
\begin{equation*}
\sim \rho^{2s}\,\int^{i\infty}dj\,j^{2s-1}\,
\left[e^{\rho(\overline{w}'-w)}\right]^j\sim (w-\overline{w}')^{-2s},
\end{equation*}
where $\rho=4/N$ for the first term in \eqref{M1J} and $\rho=2$ for the second one.
Therefore the singularity at $\Im j=\pm\infty$ cancels between
these terms and we can deform the contour as in Figure~\ref{fig:1}.

Notice that the second term in
\eqref{M1J} differs from the identity kernel \eqref{ws1} only by the
the factor $(-1)^{s+j}$. Therefore deformation of $j$
as in \eqref{j2} for the first term and as in \eqref{j1}
for the second term yields
\begin{multline}
M_{s,N}^{11}(z,\oz')=\int_{-\infty}^{\infty}d\kappa\,
e^{i\kappa(w-\ow')}\,\bigl(
\cosh w\cosh\overline{w}'
\bigr)^{2s}
\\
\times
\Bigl\{e^{\pm \pi\kappa N/4}B_{s,N}(\kappa)-e^{\pm\pi\kappa/2}
A_s(\kappa)\Bigr\},
\label{Ms11}
\end{multline}
where $B_{s,N}(\kappa)$ is \eqref{Bs} and $A_s(\kappa)$ is
\eqref{As}. The asymptotic behavior is
\begin{equation}
\sim\frac{e^{\pm N\pi\kappa/4}}{e^{N\pi|\kappa|/4}}
-\frac{e^{\pm\pi\kappa/2}}{e^{\pi|\kappa|/2}},
\label{B7.5}
\end{equation}
so the contour at infinity cancels if we choose the same sign
in both terms.

\subsection{Summary}
Here we present the results of the calculations performed in this
section. First we list the eigenvalues of operators
\eqref{Mmaps} for general $N$ and discuss their properties.
Second we present the results for $N=3$ and certain values of $s$.

\subsubsection{$N$-string Neumann eigenvalues}
\label{proposal}
The eigenfunctions have to be normalized by dividing by
$A_s(\kappa)$ from \eqref{As}, so the Neumann eigenvalues are
\begin{subequations}
\begin{align}
\mu^{II}_{s,N}(\kappa)&=e^{\pm \pi\kappa N/4}\frac{B_{s,N}(\kappa)}{A_s(\kappa)}
-e^{\pm\pi\kappa/2},
\label{MII}
\\
\mu^{IJ}_{s,N}(\kappa)&=e^{+\frac{\pi \kappa}{4}\,(N+2I-2J)}
\,\frac{B_{s,N}(\kappa)}{A_s(\kappa)}\quad (I<J),
\\
\mu^{IJ}_{s,N}(\kappa)&=(-1)^{2s}e^{-\frac{\pi \kappa}{4}\,(N-2I+2J)}
\,\frac{B_{s,N}(\kappa)}{A_s(\kappa)}\quad (I>J),
\end{align}
\label{M's}
\end{subequations}
\!\!\!where $(-1)^{2s}$ reflects the symmetry or skewsymmetry
of the Neumann matrices ($\mu^{JI}_{s,N}(\kappa)=(-1)^{2s}
\mu^{IJ}_{s,N}(-\kappa)$), and
\begin{equation}
\frac{B_{s,N}(\kappa)}{A_s(\kappa)}=\left[\frac{2}{N}\right]^{2s-1}\,
\frac{\Gamma\Bigl(s+\frac{iN\kappa}{4}\Bigr)\Gamma\Bigl(s-\frac{iN\kappa}{4}\Bigr)}
{\Gamma\Bigl(s+\frac{i\kappa}{2}\Bigr)\Gamma\Bigl(s-\frac{i\kappa}{2}\Bigr)}.
\label{B/A}
\end{equation}
The sign in \eqref{MII} is undetermined. For $s=0,1$ it makes
no difference. For $s=\frac12$ the ``$+$'' sign agrees with
other authors.
From \eqref{B/A} one easily obtains the recurrence formula relating  eigenvalues
for $s$ and $s+1$:
\begin{equation}
\frac{B_{s,N}(\kappa)}{A_s(\kappa)}=\frac{B_{s+1,N}(\kappa)}{A_{s+1}(\kappa)}\,
\frac{\kappa^2+4s^2}{\kappa^2+\frac{16}{N^2}s^2}.
\label{B/As->B/As+1}
\end{equation}
First, this shows that $B_s(\kappa)/A_s(\kappa)$
is not a continuous function at the point $(s,\kappa)=(0,0)$:
\begin{equation}
\lim_{s\to 0}\frac{B_{s,N}(0)}{A_s(0)}=\frac{N}{2}
\quad\text{while}\quad
\lim_{\kappa\to 0}\frac{B_{0,N}(\kappa)}{A_0(\kappa)}=\frac{2}{N}.
\label{N22N}
\end{equation}
This discontinuity will be important in Section~\ref{sec:zeromodes}, when
we will analyze the spectrum of $s=0$ Neumann matrices.
Second, from equation \eqref{B/As->B/As+1} it follows that if
we first take the limit $s\to 0$ then the
continuous eigenvalues \eqref{M's}  for
$s=0$ and $s=1$ coincide \cite{dima1}.

For $2s=\text{integer}$, the products of $\Gamma$-functions
reduce to hyperbolic ones. In this case \eqref{B/A} takes
the following form
\begin{equation}
\frac{B_{1,N}(\kappa)}{A_1(\kappa)}=
\frac{\sinh\frac{\pi\kappa}{2}}{\sinh\frac{N\pi\kappa}{4}}=
\frac{B_{0,N}(\kappa)}{A_0(\kappa)}
\label{B1/A1}
\end{equation}
for $s=1$ or $s=0$ and
\begin{equation}
\frac{B_{\frac12,N}(\kappa)}{A_{\frac12}(\kappa)}=
\frac{\cosh\frac{\pi\kappa}{2}}{\cosh\frac{N\pi\kappa}{4}}.
\end{equation}
for $s=\frac12$.

\subsubsection{Sliver and identity Neumann eigenvalues}

From \eqref{MII} one can easily get the eigenvalue for
the sliver Neumann matrix \cite{rastelli,okuyama1}
\begin{subequations}
\begin{multline}
\Xi_{1}(\kappa)=\lim_{N\to\infty} \mu_{1,N}^{11}(\kappa)
\\
=\lim_{N\to\infty}\left[\frac{\sinh\frac{\pi\kappa}{2}}{\sinh\frac{\pi N\kappa}{4}}\,
e^{\frac{\pi N\kappa}{4}}-e^{\frac{\pi\kappa}{2}}\right]
=-e^{-\frac{\pi|\kappa|}{2}}.
\label{xi1}
\end{multline}
Actually one can do even better. By \eqref{MII}
\begin{equation}
\Xi_s(\kappa)=\lim_{N\to\infty} \mu_{s,N}^{11}(\kappa)=
\theta(\pm\kappa)\,\frac{|\kappa|^{2s-1}}{A_s(\kappa)}
-e^{\pm\frac{\pi\kappa}{2}}.
\end{equation}
For $s=\frac12$ one obtains
\begin{equation}
\Xi_{\frac12}(\kappa)=\pm\sgn(\kappa)\,e^{-\frac{\pi|\kappa|}{2}},
\end{equation}
\end{subequations}
which agrees with \cite{arevefa} (equation (3.22))
if we choose the ``$+$'' sign in \eqref{MII}.

The identity state is a surface state \eqref{mbra} determined by the map
\eqref{h} for $N=1$. In our notation it is $\langle V_1|$.
Hence it can be represented as an exponential of  a quadratic form,
which is defined by \eqref{Mmaps} for $N=1$.
We will denote $M_{s,N=1}\equiv \Ic_s$. The spectrum
of the operator $\Ic_s(z,\oz')$ can be determined from
the general formula \eqref{M's}. For the special cases $s=0,1$ and $\frac12$
it is
\begin{equation}
\Ic_{0}(\kappa)=\Ic_{1}(\kappa)=1
\quad\text{and}\quad \Ic_{\frac12}(\kappa)=\mp\tanh\frac{\pi\kappa}{4}.
\end{equation}
The spectrum of the identity state for $s=\frac12$ agrees with that
found in \cite{arevefa} (equation (3.21)) if we choose the upper
``$-$'' sign.

\subsubsection{$3$-string Neumann eigenvalues}
Finally we specialize to $N=3$, abbreviating
\begin{equation}
\rx\equiv\frac{\pi\kappa}{4}.
\label{x}
\end{equation}
In the following formulae we suppress the index $N=3$ in
$M^{IJ}_{s,N}$.
Then for $s=1$ or $s=0$,
\begin{subequations}
\begin{equation}
\begin{split}
\mu_{1,0}^{II}(\kappa)&=-\frac{\sinh \rx}{\sinh 3\rx},
\\
\mu_{1,0}^{I,I+1}(\kappa)&=
e^{+\rx}\,
\frac{\sinh 2\rx}{\sinh 3\rx},
\\
\mu_{1,0}^{I+1,I}(\kappa)&=e^{-\rx}\,
\frac{\sinh 2\rx}{\sinh 3\rx};
\end{split}
\label{M10}
\end{equation}
and for $s=\frac{1}{2}$
\begin{equation}
\begin{split}
\mu_{\frac12}^{II}(\kappa)&=\pm\frac{\sinh \rx}{\cosh 3\rx},
\\
\mu_{\frac12}^{I,I+1}(\kappa)&=+
e^{+\rx}\,
\frac{\cosh 2\rx}{\cosh 3\rx},
\\
\mu_{\frac12}^{I+1,I}(\kappa)&=-e^{-\rx}\,
\frac{\cosh 2\rx}{\cosh 3\rx}.
\end{split}
\label{M1/2}
\end{equation}
\end{subequations}

For $s=1$, \eqref{M10} exactly coincides with Rastelli et al.
\cite{spectroscopy}. The sign ambiguity in \eqref{MII} cancels for
$s=1,0$ but not for $s=\frac{1}{2}$. For $s=\frac12$,
$\mu^{IJ}_{\frac12}(\kappa)$ in \eqref{M1/2} agrees with
Marino and Schiappa
\cite{marino} if we choose the upper ``$+$'' sign.

For $s=0$ the continuous eigenvalues are the same as for the
case $s=1$. This is in agreement with previous authors \cite{dima1}.
The improvement here is that we have much simpler expressions
for the eigenvectors as compared to \cite{dima1}.

In Section~\ref{sec:zeromodes} we will consider
the zero modes more carefully. The continuum eigenvalues
are indeed identical for $s=0$ and $s=1$.
However there is
an additional discrete state at $\kappa=0$ whose
function is to replace the average position $x$ by
the midpoint position.

\section{Zero modes, $s\to 0$ limit}
\label{sec:zeromodes}
\setcounter{equation}{0}
\subsection{The $L_0$ basis}

The correct procedure for regularizing zero modes
goes back to the earliest days of string theory \cite{fubini}.
First note that for $s\approx 0$, \eqref{id} becomes
\begin{equation*}
\frac{1}{2s}-\log\bigl(1-z\oz'\bigr)=\frac{1}{2s}+
\sum_{m=1}^{\infty}\frac{(z\oz')^m}{m}.
\end{equation*}
The divergence implements momentum conservation
\begin{equation}
\exp\left[-\frac{1}{2s}\sum_{I,J}p_Ip_J\right]\sim\delta_D\bigl(\sum_I p_I\bigr),
\label{cons}
\end{equation}
and is an essential part of the representation.

Now consider the $m=0$ oscillator. By \eqref{Lnz} it has
frequency $s\to 0$. We therefore define
\begin{equation}
a_0=\frac{1}{2}\sqrt{\frac{s}{\app}}\,x+i\sqrt{\frac{\app}{s}}\,p
\label{a0}
\end{equation}
to get
\begin{equation*}
s\left(a_0^{\dag}a_0+\frac{1}{2}\right)=
\app p^2+\frac{s^2x^2}{4\app},
\end{equation*}
which is the correct oscillator Hamiltonian.
By \eqref{basisn} $N_0^{(s)}=(2s)^{-1/2}$, so the zero
modes of the boson field become
\begin{equation}
\frac{1}{\sqrt{2s}}\,\left[\frac{\app}{2}\right]^{\frac12}\bigl(
a_0^{\dag}+a_0z^{-2s}
\bigr)=\frac{x}{2}-i\app p\,\log z,
\label{aaxp}
\end{equation}
which agrees with the usual expansion \cite{GSW}.

Next consider how the transformation $T$ \eqref{z'z} is
represented for $s\to 0$. The matrix elements $T_{mn}^{(s)}$ are defined
by
\begin{equation*}
T_{mn}^{(s)}=\langle n,s|T|m,s\rangle.
\end{equation*}
In other words
\begin{equation}
N_m^{(s)}(\bar{\beta} z+\bar{\alpha})^{-2s}\,
\left(\frac{\alpha z+\beta}{\bar{\beta} z+\bar{\alpha}}\right)^m
=\sum_{n=0}^{\infty}\,T_{mn}^{(s)}N_n^{(s)}\,z^n.
\label{C_mn}
\end{equation}
If both $m,n\geqslant 1$, the $s\to 0$ limit is nonsingular.
Differentiation with respect to $z$ of this equation for $s=0$
and comparison of the result with equation \eqref{C_mn}
for the case $s=1$ yields
\begin{subequations}
\begin{equation}
T_{mn}^{(0)}=T_{m-1,n-1}^{(1)},\quad m,n\geqslant 1.
\end{equation}
The singular cases are
\begin{align}
T_{00}^{(0)}&=1-2s\,\log \bar{\alpha},
\\
T_{m0}^{(0)}&=\sqrt{\frac{2s}{m}}\,\left(\frac{\beta}{\bar{\alpha}}\right)^m
\\
T_{0n}^{(0)}&=\sqrt{\frac{2s}{n}}\,\left(-\frac{\bar{\beta}}{\bar{\alpha}}\right)^n
\end{align}
\label{Tmn}
\end{subequations}
The $\sqrt{2s}$ zeros cancel against $ip/\sqrt{2s}$ in \eqref{a0}. The only
remaining divergence comes from the first term of $T_{00}^{(0)}$ and
enforces momentum conservation by \eqref{cons}.

\subsection{The $K_1$ basis}
In the discrete basis, the infinite norm
state $|0,0\rangle$ is clearly separated,
and the $s=0$ divergences are well defined.
In the $K_1$ basis this is not so. The spectrum is
continuous and there is an infinite norm discrete
``eigenvalue'' sitting on top of it at $\kappa=0$.
However, when we go to the second quantized world sheet
theory, these mathematical difficulties vanish.
Nevertheless we first present an heuristic first
quantized discussion, since otherwise the rigorous proof
would be hard to follow.

\subsubsection{First quantized discussion}
As is well known, the $s=1$ field $-iP(z)$ is
the derivative of the $s=0$ field $X_L(z)$.
The $L_0$ eigenfunctions \eqref{basisn}
are related by
\begin{equation}
|m+1,0\rangle(z)=\int_0^z dz' |m,1\rangle(z').
\end{equation}
Notice that the singular eigenfunction $|0,0\rangle(z)$ cannot
be obtained from $s=1$ $L_0$ eigenfunctions.
Similarly the $s=1$ eigenfunctions of $K_1$
\begin{equation}
|\kappa,1\rangle(z)=\bigl[A_1(\kappa)\bigr]^{\frac12}\,
\bigl(\cosh w\bigr)^2 e^{i\kappa w}
\end{equation}
can be integrated using
\begin{equation*}
dz=i\bigl(\cosh w\bigr)^{-2}\,dw
\end{equation*}
to give
\begin{equation}
\int_0^z dz' |\kappa,1\rangle(z')=\bigl[A_1(\kappa)\bigr]^{\frac12}\,
\frac{e^{i\kappa w}-1}{\kappa}
\equiv |\kappa,\Omega\rangle(z).
\label{kappaO}
\end{equation}
This is not quite an $s=0$ eigenfunction of $K_1$
because the $\kappa=0$ singularity has been subtracted.
The subtracted piece has no $z$ dependence and
therefore corresponds to $m=0$ in the $L_0$ basis.
This agrees with \eqref{228}. From equation \eqref{kappaO}
it follows that the function $|\kappa,\Omega\rangle(z)$
has the following expansion in terms of $s=1$
polynomials \eqref{223}
\begin{multline}
|\kappa,\Omega\rangle(z)=\sqrt{A_1(\kappa)}\,\sum_{m=1}^{\infty}
V_{m-1}^{(1)}(\kappa)\,\frac{z^m}{m}
\\
\equiv\sum_{m=1}^{\infty}\langle\kappa,1|m-1,1\rangle\,
\frac{z^m}{\sqrt{m}}.
\label{kappaRZ}
\end{multline}
Notice that the function $|\kappa,\Omega\rangle(z)$
is exactly the one found by Rastelli et al. \cite{rastelli}.

Now by \eqref{220}
\begin{equation}
A_s(\kappa)=\frac{A_{s+1}(\kappa)}{\kappa^2+4s^2}.
\label{Ass+1}
\end{equation}
As $s\to 0$ the poles at $\kappa=\pm 2is$ pinch
the $\kappa$ integral, making $\kappa=0$ very singular.
The square root in \eqref{<mk>} may have arbitrary sign.
We choose $\sqrt{A_1(\kappa)}$
always positive, and $\sqrt{A_0(\kappa)}$
to have the sign of $\kappa$.
Hence the missing piece of \eqref{kappaO} is
\begin{multline}
\lim_{s\to 0} \sqrt{A_s(\kappa)} =
\lim_{s\to 0} \frac{\sgn(\kappa)}{\sqrt{\kappa^2+4s^2}}\, \sqrt{A_{1+s}(\kappa)}
\\
=\mathscr{P}\,\frac{\sqrt{A_1(\kappa)}}{\kappa}.
\label{510}
\end{multline}
This completes the continuum wave function
\begin{equation}
\lim_{s\to 0}\,|\kappa,s\rangle(z)=
|\kappa,\Omega\rangle(z)+\mathscr{P}\,\frac{\sqrt{A_1(\kappa)}}{\kappa}.
\end{equation}

However as in \eqref{limits}
\begin{equation*}
\lim_{s\to 0}\frac{2s}{\kappa^2+4s^2}=\pi\delta(\kappa),
\end{equation*}
so there is very likely to be a discrete state at $\kappa=0$ picking
up $O(s)$ terms. The easiest way to guess its wave function $|D\rangle$
is to reverse the order of limits.
Taking $\kappa=0$ first, \eqref{kappas} and \eqref{As} yield
\begin{equation}
|\kappa=0,s\rangle=\frac{2^{s-1}}{\sqrt{\pi}}\,\Gamma(s)\bigl(
\cosh w\bigr)^{2s}\propto \frac{1}{s}+2\log\cosh w.
\label{512}
\end{equation}
Now $\cosh w=(1+z^2)^{-\frac12}$, so plausible
matrix elements with the $L_0$ basis \eqref{basisn} are
\begin{multline}
\langle 0|D\rangle \equiv 1,\quad
\langle 2j+1|D\rangle =0
\\
\text{and}\quad\langle 2j|D\rangle =(-1)^j\sqrt{\frac{2s}{2j}}
\quad\text{for } j\geqslant 1.
\label{scp}
\end{multline}
We renormalized \eqref{512} by requiring $\langle 0|D\rangle \equiv 1$.
Just as in \eqref{Tmn}, the $\sqrt{2s}$ in
\eqref{scp} can cancel against $p/\sqrt{2s}$ in \eqref{a0}.
One can also derive \eqref{scp} from the second term
in \eqref{C3'}, or by computing the residues at the pinching
poles in Figure~\ref{fig:1}, but none of these first quantized
derivations can be considered rigorous.

\subsubsection{Second quantized eigenfunctions}
Our rigorous proof will start from a different $s=0$ basis,
intermediate between $L_0$ and $K_1$. By \eqref{id} and
\eqref{218} the $s=1$ Cauchy kernel is
\begin{equation*}
\mathrm{Id}_1(z,\oz')=\bigl[1-z\oz'\bigr]^{-2}=
\int_{-\infty}^{\infty}d\kappa\,|\kappa,1\rangle(z)
\otimes \langle\kappa,1|(\oz').
\end{equation*}
Integrating both sides, we get part of the $s=0$
Cauchy kernel:
\begin{multline}
\mathrm{Id}_{s\to 0}(z,\oz')=
\Gamma(2s)-\log\bigl(1-z\oz'\bigr)+O(s)
\\
=\Gamma(2s)+\int_0^zd\zeta\,\int_0^{\oz'}d\bar{\zeta}'\,
\bigl[1-\zeta\bar{\zeta}'\bigr]^{-2}
\\
=|0,s\rangle\otimes\langle 0,s|+
\int_{-\infty}^{\infty}d\kappa\,|\kappa,\Omega\rangle(z)
\otimes \langle\kappa,\Omega|(\oz').
\label{514}
\end{multline}
Here the first outer product is the $m=0$ state from the $L_0$
basis \eqref{basisn}, and the integral is over the
outer product of \eqref{kappaO}. So if we add the $L_0$
zero mode to the continuum states \eqref{kappaO}, we get a complete
orthogonal basis for $s=0$. The divergence is all concentrated
in the first term, so it is unproblematic. Of course,
it does not quite diagonalize $K_1$, because
\eqref{510} needs to be added to \eqref{kappaO}.

Now we consider the world sheet fields. In the notations
of \cite{GSW}
\begin{equation}
X_L(z)=\frac{1}{2}\,x-i\app p\log z+i\left[\frac{\app}{2}\right]^{\frac12}
\sum_{m\ne 0}\frac{\alpha_m}{m z^m},
\label{X_L}
\end{equation}
so oscillators satisfying $[a_m^-,\,a_n^+]=\delta_{mn}$,
and occurring in $X_L(z)$ multiplied by
$s=0$ normalized $L_0$ eigenfunctions \eqref{basisn},
are
\begin{subequations}
\begin{align}
a_0^{\pm}&=\frac{1}{2}\sqrt{\frac{s}{\app}}\,x\mp i\sqrt{\frac{\app}{s}}\,p,
\quad\text{as in \eqref{a0}},
\label{516a}
\\
a_m^{\pm}&=\mp \frac{i}{\sqrt{m}}\,\alpha_{\mp m},\quad
m\geqslant 1.
\end{align}
\end{subequations}
If we differentiate \eqref{X_L}, $a_m^+$ are
multiplied by $\sqrt{m}z^{m-1}=|m-1,1\rangle(z)$,
so $a^{\pm}_{m-1}$ are also (up to a
phase factor) normalized oscillators
for $s=1$ with just a trivial index shift.
We can now put these together with the matrix elements
\eqref{<mk>} to form continuum oscillators in the $K_1$
basis. For $s=1$ the transformation is unitary, so
we get as in \eqref{phipm}
\begin{equation}
a^{\pm}(\kappa)=\mp i\sqrt{A_1(\kappa)}\,\sum_{m=1}^{\infty}
\frac{\alpha_{\pm m}}{m}\,V_{m-1}^{(1)}(\kappa),
\label{517}
\end{equation}
satisfying
\begin{equation}
[a^-(\kappa),\,a^+(\kappa')]=\delta(\kappa-\kappa').
\label{518}
\end{equation}
By \eqref{228} we only have to add an $m=0$ term to get the
$s=0$ continuum oscillators. By \eqref{<mk>}
\begin{equation*}
\langle 0,s|\kappa,s\rangle=\bigl[A_s(\kappa)\bigr]^{\frac12}\,\sqrt{2s}.
\end{equation*}
Using \eqref{510} and  multiplying by \eqref{516a}
one obtains oscillators containing the momentum $p$
\begin{equation}
a^{\pm}(\kappa,p)=\mp i\sqrt{2\app}\, \hat{p}\,\mathscr{P}\frac{\sqrt{A_1(\kappa)}}{\kappa}
+a^{\pm}(\kappa).
\label{519}
\end{equation}
The undefined term $\bigl(\mathscr{P}\,\frac{1}{\kappa}\bigr)^2$ cancels
from the commutator, so these are satisfactory oscillators in the
$K_1$ basis.

Now we recall \eqref{514}. The first outer product is the zero mode from
the $L_0$ basis. It corresponds to the oscillator
$a_0^{\pm}$ of \eqref{516a}. The second outer
product is the integrated $s=1$ $K_1$ basis \eqref{kappaO}. It
corresponds to the oscillators $a^{\pm}(\kappa)$ of
\eqref{517}. The $z$ integration just provides the $1/m$.
So in this basis we can take the $s\to 0$ limit simply by replacing $a_0^{\pm}$
by $\hat{x}$ and $\hat{p}$. Thus \eqref{514} implies that
\begin{equation}
\hat{x},\,\hat{p}\quad\text{and}\quad a^{\pm}(\kappa)
\end{equation}
form a complete orthogonal basis for the $s=0$ world sheet
field theory. We need to replace $a^{\pm}(\kappa)$ by
$a^{\pm}(\kappa,p)$, but the extra term in \eqref{519}
can be added by a unitary transformation. Define
\begin{multline}
U_p=\exp\biggl\{i\sqrt{2\app}\,\hat{p}
\int_{-\infty}^{\infty}d\kappa\,
\mathscr{P}\frac{\sqrt{A_1(\kappa)}}{\kappa}\,
\\
\times
\Bigl[a^{+}(\kappa)+a^{-}(\kappa)\Bigr]\biggr\}.
\label{U}
\end{multline}
Then by \eqref{518}
\begin{subequations}
\begin{align}
a^{\pm}(\kappa,p)&=U^{-1}_p\,a^{\pm}(\kappa)\,U_p.
\label{523a}
\\
\intertext{Although the momentum operator does not change
under this unitary transformation}
\hat{p}&=U^{-1}_p\, \hat{p}\, U_p,
\\
\intertext{the center of mass coordinate operator changes to}
\hat{\xi}&\equiv U^{-1}_p\,\hat{x}\, U_p.
\end{align}
\end{subequations}
To calculate $\hat{\xi}$, we insert \eqref{517} into
\eqref{U} and do the integral by \eqref{C5}, getting
\begin{equation}
U_p=\exp\left[\sqrt{2\app}\,\hat{p}
\sum_{m=1}^{\infty}\frac{(-1)^m}{2m}\bigl(
\alpha_{2m}-\alpha_{-2m}\bigr)\right].
\label{Ud}
\end{equation}
Then by \eqref{X_L}
\begin{multline}
\hat{\xi}=\hat{x}+i\sqrt{2\app}\,
\sum_{m=1}^{\infty}\frac{(-1)^m}{2m}\bigl(
\alpha_{2m}-\alpha_{-2m}\bigr)
\\
\equiv X_L(i)+X_L(-i).
\label{xi}
\end{multline}
This is just the position of the string's midpoint.
\eqref{xi} also confirms the guess \eqref{scp}.
The oscillator corresponding to the discrete $\kappa=0$
state is
\begin{equation*}
a_D^{\pm}=\frac{1}{2}\sqrt{\frac{s}{\app}}\,\xi\mp i\sqrt{\frac{\app}{s}}\,p,
\end{equation*}
in analogy to \eqref{516a}.

The unitary transformation \eqref{U} cannot change
the commutators, so we conclude finally that $\hat{\xi},\,\hat{p}$ and
$a^{\pm}(\kappa,p)$ form a complete orthogonal basis for
the $s=0$ world sheet field theory with
$K_1$ diagonal, and
\begin{equation}
\begin{split}
&[\hat{\xi},\,\hat{p}]=i,\qquad
[a^{-}(\kappa,p),\,a^+(\kappa',p)]=\delta(\kappa -\kappa'),
\\
&
[\hat{\xi},\,a^{\pm}(\kappa,p)]=
[\hat{p},\,a^{\pm}(\kappa,p)]=0.
\end{split}
\end{equation}

The average  position $x$ has been replaced by the midpoint position
$\xi$. In view of the importance of the string midpoint in
Witten's string field theory, this is a very satisfying result.

\subsubsection{Expansion of the world sheet field
in $K_1$ eigenfunctions}
Lastly we expand $X_L(z)$ in these oscillators, assuming
that $z$ is on the boundary of the unit disk.
Again we start with the basis \eqref{514}, where
the wave functions \eqref{kappaO} give by \eqref{kappaRZ}
\begin{multline}
X_L(z)=
\frac{1}{2}\,x-i\app p\log z
+\left[\frac{\app}{2}\right]^{\frac12}\times
\\
\times
\int_{-\infty}^{\infty}d\kappa\,
\Bigl\{a^+(\kappa)\,|\kappa,\Omega\rangle(z)+
a^-(\kappa)\,|\kappa,\Omega\rangle(\oz)\Bigr\}
\end{multline}
By \eqref{C1}
\begin{multline*}
\lim_{s\to 0}\int_{-\infty}^{\infty}d\kappa\,A_s(\kappa)\,
\bigl[e^{i\kappa w}-1\bigr]
\\
=\lim_{s\to 0}\Gamma(2s)
\Bigl[\bigl(\cosh w\bigr)^{-2s}-1\Bigr]
=-\log\cosh w
\end{multline*}
or by \eqref{kappaO} and \eqref{510}
\begin{multline}
-\frac12\,\log(1+z^2)=\log\cosh w
\\
=-\int_{-\infty}^{\infty}d\kappa\,
\mathscr{P}\frac{\sqrt{A_1(\kappa)}}{\kappa}\,
|\kappa,\Omega\rangle(z).
\label{logcosh}
\end{multline}
For $z$ on the unit circle, $\log z=\frac12\,\log\left(
\frac{1+z^2}{1+\oz^2}\right)$, so by \eqref{519}
\begin{multline*}
X_L(z)=
\frac{1}{2}\,x+\left[\frac{\app}{2}\right]^{\frac12}\times
\\
\times
\int_{-\infty}^{\infty}d\kappa\,
\Bigl\{a^+(\kappa,p)\,|\kappa,\Omega\rangle(z)+
a^-(\kappa,p)\,|\kappa,\Omega\rangle(\oz)\Bigr\}.
\end{multline*}
Notice that the oscillators were changed from $a^{\pm}(\kappa)$
to $a^{\pm}(\kappa,p)$. The next step is to change $x$ to $\xi$.
To this end we substitute expression \eqref{kappaO} for the function
$|\kappa,\Omega\rangle(z)$ and split it into two terms by
changing $\frac{1}{\kappa}$ to $\mathscr{P}\frac{1}{\kappa}$.
\begin{multline*}
X_L(z)=
\frac{1}{2}\,x+\left[\frac{\app}{2}\right]^{\frac12}
\int_{-\infty}^{\infty}d\kappa\,\mathscr{P}\,
\frac{\sqrt{A_1(\kappa)}}{\kappa}
\\
\times
\Bigl\{a^+(\kappa,p)\,e^{i\kappa w}+
a^-(\kappa,p)\,e^{-i\kappa \ow}
\\
-\bigl[a^+(\kappa,p)+a^-(\kappa,p)\bigr]\Bigr\}.
\end{multline*}
Now notice that the last term in the integral can
also be written by \eqref{519} as $a^{+}(\kappa)+a^{-}(\kappa)$,
i.e. the terms proportional to the momentum cancel.
Hence the last term can be integrated as in \eqref{U}~$\to$~
\eqref{Ud}. Finally one obtains
\begin{multline}
X_L(z)=\frac{1}{2}\,\xi
+\left[\frac{\app}{2}\right]^{\frac12}
\int_{-\infty}^{\infty}d\kappa\,\mathscr{P}\,
\frac{\sqrt{A_1(\kappa)}}{\kappa}
\\
\times
\Bigl\{a^+(\kappa,p)\,e^{i\kappa w}+
a^-(\kappa,p)\,e^{-i\kappa \ow}\Bigr\}.
\label{X_Lcont}
\end{multline}
This equation contains one subtlety: because
of the singularity in the oscillators $a^{\pm}(\kappa,p)$
the integral cannot be rewritten as a sum of two integrals
corresponding to creation and annihilation parts.
We emphasize that it is only valid for $\Im w=\pm\frac{\pi}{4}$,
corresponding to $|z|=1$.

\section{$s=0$ Neumann matrices}
\label{sec:zeroNeumann}
\setcounter{equation}{0}
\subsection{Zero mode vertex}
In this section we apply these zero mode fields to calculate the
$s=0$ Neumann matrix.

Let $I,J=1,\dots,N$ label the external lines. Then the $s=1$
vertex in the diagonal $K_1$ basis is
\begin{widetext}
\begin{equation}
\langle V_N^{(0)}|={}_{1\dots N}\langle 0|\exp\left\{
\frac{1}{2}\,\sum_{I,J=1}^{N}\int_{-\infty}^{\infty}
d\kappa\,a^{-(I)}(\kappa)\,\mu^{IJ}_{N}(\kappa)\,\bigl(Ca^{-(J)}\bigr)(\kappa)
\right\}.
\end{equation}
Here $a^{-(I)}(\kappa)$ are the $s=1$ annihilation oscillators
\eqref{517} acting in the $I$-th particle Hilbert space,
$\mu_N^{IJ}(\kappa)$ are the eigenvalues \eqref{M's}
for $s=1$ or $s=0$ and $C$ is a twist operator, which
acts on the oscillators $a^{-}(\kappa)$ as
\begin{equation}
\bigl(C a^{\mp}\bigr)(\kappa)=-a^{\mp}(-\kappa).
\end{equation}
We now include the momenta by
applying $N$ copies of the unitary transformation
\eqref{U}. By \eqref{Ass+1} this can be written
\begin{equation}
U_{p^I}=\exp\biggl\{
i\sqrt{2\app}\,p^I\,\int_{-\infty}^{\infty}
d\kappa\,\lim_{s\to 0}\sqrt{A_s(\kappa)}\,
\bigl[
a^{+(I)}(\kappa)+a^{-(I)}(\kappa)
\bigr]
\biggr\}
\end{equation}
which can be normal ordered by \eqref{C1}
\begin{equation}
U_{p^I}=\exp\left\{
-\app \bigl(p^I\bigr)^2\,\Gamma(2s)
\right\}\,:U_{p^I}:\,.
\end{equation}
Notice now that $\sqrt{A_s(\kappa)}$ contains
$\sgn(\kappa)$ in its definition \eqref{510} and therefore
it is even with respect to action of the twist operator $C$.
Thus by \eqref{523a} and \eqref{519}
\begin{multline}
\langle V_N^{(0)}|\{p^I\}\rangle
\equiv\langle V_N^{(0)}|\bigotimes_{I=1}^N\,U_{p^I}
=\lim_{s\to 0}\,{}_{1\dots N}\langle 0|\exp\left\{-\Gamma(2s)\,\sum_{I=1}^N
\app\bigl(p^I\bigr)^2
\right.
\\
+\frac{1}{2}\,\sum_{I,J=1}^N\,\int_{-\infty}^{\infty}d\kappa\,
a^{-(I)}(\kappa,p)\,\mu^{IJ}_{N}(\kappa)\,\bigl(C a^{-(J)}\bigr)(\kappa,p)
\left.
+i\sqrt{2\app}\,\sum_{I=1}^N\,
p^I\int_{-\infty}^{\infty}d\kappa\,\sqrt{A_s(\kappa)}\,
a^{-(I)}(\kappa)
\right\}
\end{multline}
The creation part of $U_{p^I}$ converted $a^{-(I)}(\kappa)$
to $a^{-(I)}(\kappa,p)$ as in \eqref{523a},
while the annihilation part gave an extra diagonal piece.
The term in the first line and the integral in the second line
are singular. To show that the expression in the exponent is
meaningful let us rewrite it in terms of oscillators $a^{-}(\kappa)$:
\begin{multline}
\langle V_N^{(0)}|\{p^I\}\rangle
=\lim_{s\to 0}\,{}_{1\dots N}\langle 0|\exp\left\{
+\frac{1}{2}\,\sum_{I,J=1}^N\,\int_{-\infty}^{\infty}d\kappa\,
a^{-(I)}(\kappa)\,\mu^{IJ}_{N}(\kappa)\,\bigl(C a^{-(J)}\bigr)(\kappa)
\right.
\\
+i\sqrt{2\app}\,\sum_{I=1}^N\,
p^I\int_{-\infty}^{\infty}d\kappa\,
\sqrt{A_s(\kappa)}\,\bigl[\mu^{IJ}_N(\kappa)+\delta^{IJ}
\bigr]\,\bigl(C
a^{-(I)}\bigr)(\kappa)
\left.
-\app\sum_{I,J=1}^{N} p^I\,p^J\,\int_{-\infty}^{\infty}d\kappa\,A_s(\kappa)
\bigl[\mu^{IJ}_N(\kappa)+\delta^{IJ}\bigr]
\right\}
\label{VN:1}
\end{multline}
\end{widetext}
We see now that the integral in the first line is well defined,
however there are problems with the $s\to 0$ limit in the other
two integrals.

Consider first the last term in \eqref{VN:1}.
Notice the following integral of \eqref{Bs} analogous
to \eqref{C1}:
\begin{equation}
\int_{-\infty}^{\infty}d\kappa\,e^{\kappa y}\,B_{s,N}(\kappa)
=\left[\frac{2}{N}\right]^{2s}\,\Gamma(2s)
\left[\cos\Bigl(\frac{2y}{N}\Bigr)\right]^{-2s}.
\label{ABs}
\end{equation}
For $I\ne J$ the integral in the last line of \eqref{VN:1}
can be calculated by using
\eqref{M's} and \eqref{ABs} with the result
\begin{equation*}
\int_{-\infty}^{\infty}d\kappa\,A_s(\kappa)\,
\mu^{IJ}_N(\kappa)=\Gamma(2s)-M_{N,\,00}^{\prime\,IJ}+O(s)\quad(I\ne J),
\end{equation*}
where
\begin{equation}
M_{N,\,00}^{\prime\,IJ}=\bigl(1-\delta^{IJ}\bigr)\,
\log\Bigl[\frac{N}{2}\,\sin\frac{\pi}{N}|I-J|\Bigr].
\label{M'IJ00}
\end{equation}
For $I=J$ one has to insert extra regularization
by multiplying by $e^{\pm\varepsilon\kappa}$.
Once again the integral can be calculated by using \eqref{ABs},
and \eqref{C1} for the subtraction term in \eqref{MII}
\begin{multline*}
\int_{-\infty}^{\infty}d\kappa\,A_s(\kappa)\,
\mu^{II}_N(\kappa)
=\lim_{\varepsilon \to 0}
\Biggl\{
\left[\frac{N}{2}\,\sin
\frac{2\varepsilon}{N}
\right]^{-2s}
\\
-\bigl(\sin\varepsilon
\bigr)^{-2s}
\Biggr\}
=0\quad
\text{for}\quad s<1.
\end{multline*}
Hence the last term in the exponent is
\begin{equation}
-\Gamma(2s)\,\app\left(\sum p^I\right)^2
+\app \sum_{I,J=1}^{N} p^I M^{\prime\,IJ}_{N,\,00}\,p^J.
\end{equation}
The first term here is infinite. This is responsible for
momentum conservation and after proper normalization of the vertex
it yields $(2\pi)^{26}\delta(\sum p^I)$ as in \eqref{cons}.

Now we are ready to consider the second term in the
exponent of \eqref{VN:1}. Because of momentum conservation, we are at liberty to
include an extra factor
\begin{equation}
\exp\left[\Lambda\,i\sqrt{2\app}\,
\sum_{I,J=1}^{N}
p^I\int_{-\infty}^{\infty}d\kappa\,\sqrt{A_s(\kappa)}\,
a^{-(J)}(\kappa)\right],
\end{equation}
where $\Lambda$ is an arbitrary constant. We choose
$\Lambda=-2/N$, since by \eqref{N22N} and \eqref{M's}
\begin{equation}
\mu^{IJ}_N(0)=\frac{2}{N}-\delta^{IJ}.
\end{equation}
This turns $\delta^{IJ}$ in \eqref{VN:1} into $-\mu^{IJ}_N(0)$.
After this substitution the $s\to 0$ limit is easy.
Finally the $s=0$ $N$-string vertex in the $K_1$
diagonal basis takes the form
\begin{widetext}
\begin{multline}
\langle V_N^{(0)}|\{p^I\}\rangle
=(2\pi)^{26}\,\delta(p^{1}+\dots+p^{N})
\;{}_{1\dots N}\langle \Omega|\exp\left\{
+\app \sum_{I,J=1}^{N} p^I M^{\prime\,IJ}_{N,\,00}\,p^J
\right.
\\
+i\sqrt{2\app}\,\sum_{I=1}^N\,
p^I\int_{-\infty}^{\infty}d\kappa\,
\mathscr{P}\,\frac{\sqrt{A_1(\kappa)}}{\kappa}\,
\bigl[\mu^{IJ}_N(\kappa)-\mu^{IJ}_N(0)\bigr]\,\bigl(C
a^{-(I)}\bigr)(\kappa)
\\
\left.
+\frac{1}{2}\,\sum_{I,J=1}^N\,\int_{-\infty}^{\infty}d\kappa\,
a^{-(I)}(\kappa)\,\mu^{IJ}_{N}(\kappa)\,\bigl(C a^{-(J)}\bigr)(\kappa)
\right\},
\label{VN:CF}
\end{multline}
where ${}_{1\dots N}\langle \Omega|$ is a tensor
product of $N$ Fock vacua $\langle \Omega|$ for the
oscillators $a^{-}(\kappa)$,
and $M^{\prime\,IJ}_{N,\,00}$ is \eqref{M'IJ00}.
Notice that in principal one can omit $\mathscr{P}$ in the second
line of \eqref{VN:CF}.

For $N=3$ the fact that the zero and nonzero momentum matter
vertices are related by a unitary transformation
agrees with \cite{DK}.

\subsection{Vertex in the $L_0$ basis}
To compare the vertex \eqref{VN:CF} with
\cite{GJ} and \cite{peskin1}
we first have to rewrite it in the discrete $L_0$
basis. Substituting the continuum oscillators
\eqref{517} and \eqref{<mk>} we obtain the following
expression for the $N$-string
vertex \eqref{mbra} in the momentum representation:
\begin{multline}
\langle V_N^{(0)}|\{p^{(I)}\}\rangle=(2\pi)^{26}
\delta^{(26)}\bigl(p^{(1)}+\dots+p^{(N)}\bigr)\,
{}_{1\dots N}\langle \Omega|\,
\exp
\left[+\app \sum_{I,J=1}^N p^{(I)}p^{(J)}M_{N,\,00}^{\prime\,IJ}
\right.
\\
\left.
+\sqrt{2\app}\sum_{I,J=1}^{N}
\sum_{m=1}^{\infty}p^{(I)}M_{N,\,0m}^{\prime\,IJ}(-1)^m a_m^{(J)}
+\frac{1}{2}
\sum_{I,J=1}^N\sum_{m,n=1}^{\infty}
a_m^{(I)}M_{N,\,mn}^{\prime\,IJ}(-1)^n\,a_n^{(J)}
 \right],
\label{V_Nmoment}
\end{multline}
\end{widetext}
where $a_n=\frac{\alpha_n}{\sqrt{n}}$, and
${}_{1\dots N}\langle \Omega|$ is a tensor
product of $N$ Fock vacua $\langle \Omega|$ for the
oscillators $a_n^{(I)}$ ($n\geqslant 1$), and
the matrix $M_N^{\prime\,IJ}$ is defined by
\begin{subequations}
\begin{align}
M_{N,\,mn}^{\prime\,IJ}&=
-\int_{-\infty}^{\infty}d\kappa\,\mu^{IJ}_{1,N}(\kappa)\,
\langle m-1,1|\kappa,1\rangle
\notag
\\
&\quad\times\langle\kappa,1|n-1,1\rangle
\quad\text{for}\quad m,n\geqslant 1;
\label{M':mn}
\\
M_{N,\,0m}^{\prime\,IJ}&=-\int_{-\infty}^{\infty}d\kappa\,
\frac{\sqrt{A_1(\kappa)}}{\kappa}\,\Bigl[\mu_{1,N}^{IJ}(\kappa)-\mu_{1,N}^{IJ}(0)
\Bigr]
\notag
\\
&\quad\times\,
\langle\kappa,1|m-1,1\rangle
\quad\text{for}\quad m\geqslant 1;
\label{M':0m}
\\
M_{N,\,00}^{\prime\,IJ}&=\bigl(1-\delta^{IJ}\bigr)\,
\log\Bigl[\frac{N}{2}\,\sin\frac{\pi}{N}|I-J|\Bigr].
\label{N:M'00}
\end{align}
\label{M''}
\end{subequations}
The sign in equation \eqref{M':mn} comes from
the ``$i$'' in the definition of the continuum oscillator
\eqref{517}.
Notice that equation \eqref{M':mn} up to sign
and obvious shift of indexes coincides
with the $s=1$ Neumann matrix.

For $N=3$ the representation \eqref{M''} of the
momentum Neumann matrices in the $\kappa$-basis
coincides with that in \cite{DK} (see Appendix~B therein).

\subsection{Check of Neumann matrix elements}
Lastly we check the $s=0$ Neumann matrices
$M^{\prime\,IJ}_{N,\,mn}$ against \cite{GJ} and \cite{peskin1}.
Let us start from $m=n=0$.
For $N=3$ equation \eqref{N:M'00} yields
\begin{equation}
M_{00}^{\prime\,IJ}=\bigl(1-\delta^{IJ}\bigr)\log\frac{3\sqrt{3}}{4},
\end{equation}
which is in complete agrement with \cite{GJ} (paper 1).
For general $N$ equation \eqref{N:M'00}
gives the $00$ matrix element, which
after taking into account momentum conservation
coincides with the result of \cite{peskin1}
(paper 1, equation (4.27)).

Next we check $M^{\prime\,IJ}_{N,\,0m}$.
From equation \eqref{M':0m} and \eqref{<mk>} one obtains the following
integral representation
\begin{subequations}
\begin{multline}
M_{N,\,0m}^{\prime\,IJ}=-\frac{1}{\sqrt{m}}\int_{-\infty}^{\infty}d\kappa\,
\mu_{1,N}^{IJ}(\kappa)\,\mathscr{P}\,\frac{A_1(\kappa)}{\kappa}\,
V_{m-1}^{(1)}(\kappa)
\\
+\mu_{1,N}^{IJ}(0)\,\xi_m,
\label{int1}
\end{multline}
where $\xi_m$ is by \eqref{C5}
\begin{multline}
\xi_m\equiv\int_{-\infty}^{\infty}d\kappa\,\mathscr{P}\,\frac{\sqrt{A_1(\kappa)}}{\kappa}\,
\langle\kappa,1|m-1,1\rangle
\\
=\left\{%
\begin{array}{ll}
    -\frac{(-1)^k}{\sqrt{2k}}, & \hbox{for } m=2k;\\
    0, & \hbox{for } m=2k-1 \\
\end{array}%
\right.
\label{xim}
\end{multline}
\label{M'N}
\end{subequations}
and corresponds to the discrete state \eqref{scp}.
Since $V_{m-1}(\kappa)$ are polynomials
in $\kappa$, the integral in \eqref{int1} can be calculated by
differentiating the following function (see \eqref{M's})
\begin{equation}
\int_{-\infty}^{\infty}d\kappa\,\mathscr{P}\frac{B_{1,N}(\kappa)}{\kappa}\,
e^{\kappa y}=\frac{2}{N}\,\tan \frac{2y}{N},
\label{BN2}
\end{equation}
which follows from the $s\to 0$ limit \eqref{limits}
of the derivative of \eqref{ABs} with respect to $y$.
This trick nicely works when $I\ne J$. For $I=J$ one
has to insert extra regularization by multiplying
the integrand in \eqref{int1} by $e^{\varepsilon \kappa}$,
then one can calculate the integral via \eqref{BN2}
and  take the limit $\varepsilon \to 0$
after the subtraction in \eqref{MII}.

For $N=3$ the equation \eqref{M'N} gives
(we suppress index $N=3$)
\begin{subequations}
\begin{align}
\begin{split}
\frac{1}{\sqrt{2}}\,M_{02}^{\prime\,12,13}&=-\frac{8}{27}+\frac{1}{3}=\frac{1}{27},
\\
\frac{1}{\sqrt{2}}\,M_{02}^{\prime\,11}&=\frac{5}{54}-\frac{1}{6}=-\frac{2}{27};
\end{split}
\\
\begin{split}
\frac{1}{\sqrt{4}}\,M_{04}^{\prime\,12,13}&=\frac{112}{729}-\frac{1}{6}=-\frac{19}{1458},
\\
\frac{1}{\sqrt{4}}\,M_{04}^{\prime\,11}&=-\frac{167}{2916}+\frac{1}{12}=\frac{19}{729}.
\end{split}
\end{align}
\end{subequations}
Here the first two numbers are from the corresponding
terms in \eqref{M'N}, while the third is from \cite{GJ} (paper 1,
equation (4.25)).
We also checked
$mn=01,\,03,\,11,\,12,\,22$. The usual Neumann
matrices are twisted \cite{spectroscopy}, meaning that
elements with the first index odd change sign.
Allowing for this, we found complete agreement.
\bigskip

\section{Ghost vertex}
\label{sec:ghost}
\setcounter{equation}{0}
\subsection{Review of ghost gluing vertices}
The ghost gluing vertex $\langle V_N^{s,1-s}|$
is a multilinear map from the $N$-th tensor power of the
ghost Fock space to complex numbers. For a free
ghost conformal field theory the ghost vertex can be
written as a Gaussian state. However there are
two subtleties. The first one is related to
non-zero background charge $Q$ of the ghost systems
($Q=-3$ for $bc$-ghosts and $Q=2$ for $\beta\gamma$-ghosts).
The second subtlety is related to the choice of
picture but it is important only for fermionic $\beta\gamma$-ghosts,
which are bosons. The ghosts occur in conjugate
pairs with scale dimensions $s$ and $1-s$.
By \eqref{NN} $N_m^{(s)}$ cancels from the
ghost (anti)commutators if we expand
as in \eqref{32}.

\subsubsection{Ghost gluing vertex}
There are several equivalent representations for the
$bc$-ghost gluing vertex. For our purpose it
is convenient to use the representation constructed
by Gross and Jevicki \cite{GJ} (the second paper).
The other formulation can be found in \cite{peskin1}.
So the $3$-string gluing vertex is of the form \cite{GJ}
\begin{widetext}
\begin{equation}
\langle V_3^{2,-1}|=N_{\textsf{gh}}\,{}_{123}\langle +|
\exp\left[-\sum_{I,J=1}^3\sum_{m=1,n=1}^{\infty}
b_m^{(I)} \frac{1}{\sqrt{m}}\,\bigl(M^{IJ}_{bc}C\bigr)_{mn}
\sqrt{n}\,
c_n^{(J)}-\sum_{I,J=1}^3\sum_{n=1}^{\infty}
b_0^{(I)} \bigl(M^{IJ}_{bc}C\bigr)_{0n}
\sqrt{n}\,
c_n^{(J)}\right],
\label{Vgh1}
\end{equation}
where $N_{\textsf{gh}}=\left(\frac{3\sqrt{3}}{4}\right)^3$ is
a normalization constant,
${}_{123}\langle +|$ denotes the tensor
product of three $\langle +|$ vacua
\end{widetext}
from each
Fock space, $\langle +|$ is related to the $SL(2,\Rh)$
vacuum via $\langle +|=\langle 0| c_0 c_{-1}$,
$b_m^{(I)}$ and $c_n^{(I)}$ are annihilation
operators acting in the $I$-th Fock space, $C_{mn}=(-1)^m\delta_{mn}$ is a twist
operator
and $M^{IJ}_{bc}$ are the ghost Neumann matrices which we are going
to diagonalize.

There are two ways to express the operator $M^{IJ}_{bc}$
in terms of the maps $h_I$. We use one which was described
by Gross and Jevicki \cite{GJ} (paper 2). They related
the $3$-string ghost Neumann matrix $\bigl(M_{bc}^{IJ}\bigr)_{mn}$ to
the $6$-string matter ($s=0$) Neumann matrix $M^{\prime\,IJ}_{6,\,mn}$ as
\begin{equation}
M^{IJ}_{bc}=(-1)^{I+J}\bigl(-M^{\prime\,IJ}_{6}
+M^{\prime\,I,J+3}_{6}\bigr)
\label{M6->gM3}
\end{equation}
for $I,J=1,2,3$.
Here operators $M^{\prime\,IJ}_{6}$ are given by
equation \eqref{M''}. From these $6$-string Neumann matrices
one can also obtain $3$-string matter matrices by
\begin{equation}
M^{\prime\,IJ}_{3}=M^{\prime\,IJ}_{6}
+M^{\prime\,I,J+3}_{6},\quad
I,J=1,2,3.
\label{M6->M3}
\end{equation}

The second method is related to conformal gluing \cite{peskin1}
and its application will be considered elsewhere \cite{dima3}.

\subsubsection{Superghost gluing vertex}
The case of $NS$ superghosts is more complicated because
of the pictures. Here we will consider the $3$-string vertex
over the picture $-1$ vacuum \cite{GJ} (paper 3):
\begin{multline}
\langle V_3^{\,\text{sgh}}|={}_{123}\langle -1|
\\
\times\exp\left[
\sum_{I,J=1}^{3}\sum_{r,s=\frac12}^{\infty}
\beta_{r}^{(I)}\bigl(M^{IJ}_{\beta\gamma}C\bigr)_{rs}\gamma_s^{(J)}
\right]
\Delta\Bigl(\frac{\pi}{2}\Bigr),
\label{Vsgh}
\end{multline}
where ${}_{123}\langle-1|$ is a tensor product of three
Fock vacua in the $-1$ picture, $\beta_r^{(I)},\,\gamma^{(I)}_s$
are ghost/antighost annihilation operators
acting in the $I$th Fock space, $M^{IJ}_{\beta\gamma}$ are
Neumann matrices, which we are going to diagonalize,
$C_{rs}=(-1)^{r-\frac12}\delta_{rs}$ is a twist matrix
and $\Delta(\frac{\pi}{2})$ is the midpoint insertion.
The Neumann  matrix $M^{IJ}_{\beta\gamma}$ is given by the generating
function \cite{GJ} (paper 3, equation (4.35)), which in our notations takes the form
\begin{widetext}
\begin{multline}
M_{\beta\gamma}^{IJ}(z,\oz')\equiv \sum_{r,s=\frac12}^{\infty}
\bigl(M_{\beta\gamma}^{IJ}\bigr)_{rs}z^{r-\frac12}(-\oz')^{s-\frac12}
=-\frac{\delta^{IJ}}{z+\oz'}
\\
+\frac{1}{2}\left[
\left(\frac{h_I(z)}{h_J(-\oz')}\right)^{\frac12}
+
\left(\frac{h_I(z)}{h_J(-\oz')}\right)^{-\frac12}
\right]
\frac{\bigl[h_I'(z)\bigr]^{1/2}\bigl[h_J'(-\oz')\bigr]^{1/2}}{h_I(z)-h_J(-\oz')}
.
\label{superN}
\end{multline}
\end{widetext}

\subsection{$bc$-ghost $3$-string vertex}
In Section~\ref{sec:zeroNeumann} we obtained expressions \eqref{M''} for
the elements of momentum $s=0$ $N$-string Neumann matrices,
in terms of integrals
involving eigenvalues $\mu_{1,N}^{IJ}(\kappa)$. From
\eqref{M's} follows that for
$N=6$ these eigenvalues are
\begin{subequations}
\begin{align}
\mu^{II}_{1,6}(\kappa)&=-\frac{\sinh 4\rx}{\sinh 6\rx},
\label{M6II}
\\
\mu^{IJ}_{1,6}(\kappa)&=e^{2\rx(3+I-J)}
\,\frac{\sinh 2\rx}{\sinh 6\rx}\quad\text{for}\quad I<J,
\\
\mu^{IJ}_{1,6}(\kappa)&=e^{2\rx(3-I+J)}
\,\frac{\sinh 2\rx}{\sinh 6\rx}\quad\text{for}\quad I>J,
\end{align}
\label{M6's}
\end{subequations}
where $\rx\equiv\frac{\pi\kappa}{4}$.
Hence \eqref{M6->gM3} and \eqref{M''} yield the following
representation for the $3$-string ghost Neumann matrices
($m,n\geqslant 1$):
\begin{subequations}
\begin{align}
\bigl(M^{IJ}_{bc}\bigr)_{mn}&=
-\int_{-\infty}^{\infty}d\kappa\,\mu_{bc}^{IJ}(\kappa)\,
\langle m,0|\kappa,\Omega\rangle\,
\notag
\\
& \qquad\quad\qquad\qquad\times\langle \kappa,\Omega|n,0\rangle\, ,
\\
\bigl(M_{bc}^{IJ}\bigr)_{0n}&=
-\int_{-\infty}^{\infty}d\kappa\,
\bigl[\mu_{bc}^{IJ}(\kappa)-\mu_{bc}^{IJ}(0)\bigr]\,
\notag
\\
& \qquad
\times
\mathscr{P}\,\frac{\sqrt{A_1(\kappa)}}{\kappa}\,\langle \kappa,\Omega|n,0\rangle\, ,
\end{align}
\label{tM}
\end{subequations}
where the eigenvalues of the ghost Neumann matrices
$\mu_{bc}^{IJ}(\kappa)$ are given by
\begin{subequations}
\begin{align}
\mu_{bc}^{11}(\kappa) &= -\mu^{11}_{1,6}(\kappa)
+\mu^{14}_{1,6}(\kappa)=\frac{\cosh\rx}{\cosh 3\rx},
\\
\mu_{bc}^{12}(\kappa)&=+\mu^{12}_{1,6}(\kappa)
-\mu^{15}_{1,6}(\kappa)=+e^{+\rx}\,
\frac{\sinh 2\rx}{\cosh 3\rx},
\\
\mu_{bc}^{13}(\kappa)&=-\mu^{13}_{1,6}(\kappa)
+\mu^{16}_{1,6}(\kappa)=-e^{-\rx}\,\frac{\sinh
2\rx}{\cosh 3\rx}.
\end{align}
\label{gh}
\end{subequations}
These eigenvalues agree with those found in \cite{dima1,Erler},
and the continuum representation (\ref{tM}b) for
$\bigl(M_{bc}^{IJ}\bigr)_{0n}$ coincides with that in \cite{dima1}.
In addition one obtains the following relation between eigenvalues
of $3$-string matter boson Neumann matrices \eqref{M10}
and $bc$-Neumann matrices \eqref{gh}
\begin{equation}
\mu^{IJ}_{bc}(\kappa)=\mu^{IJ}_{1,3}(\kappa\pm 2i).
\label{ghm}
\end{equation}
As another check of our result one
can easily show that
the sum \eqref{M6->M3} of $6$-string Neumann
matrices \eqref{M''} indeed yields
the $3$-string matrices \eqref{M''}.
In particular, the sum \eqref{M6->M3} of
$6$-string Neumann matrix eigenvalues \eqref{M6's}
yields the $3$-string eigenvalues \eqref{M10}.

Now we will rewrite the ghost $3$-string vertex
\eqref{Vgh1} in the diagonal basis. To this end we
introduce ghost continuum oscillators:
\begin{subequations}
\begin{align}
b^{\pm}(\kappa)&=\sum_{m=1}^{\infty}b_{\mp m}\,\frac{1}{\sqrt{m}}\,
\langle m,0|\kappa,\Omega\rangle;
\\
c^{\pm}(\kappa)&=\sum_{m=1}^{\infty}c_{\mp m}\,\sqrt{m}\,
\langle m,0|\kappa,\Omega\rangle
\label{610b}
\end{align}
\label{610}
\end{subequations}
with the commutation relations
\begin{equation}
\{b^{\pm}(\kappa),\,c^{\mp}(\kappa)\}=\delta(\kappa-\kappa').
\end{equation}
The twist operator $C$ acts on the continuum oscillators as
\begin{equation*}
\bigl(C\,c^{\pm}\bigr)(\kappa)=-c^{\pm}(-\kappa)\quad
\text{and}\quad
\bigl(C\,b^{\pm}\bigr)(\kappa)=-b^{\pm}(-\kappa).
\end{equation*}
Then the vertex \eqref{Vgh1} becomes
\begin{widetext}
\begin{multline}
\langle V_3^{2,-1}|=N_{\textsf{gh}}\,{}_{123}\langle +|
\exp\biggl[
\int_{-\infty}^{\infty} d\kappa\,
b^{-(I)}(\kappa)\,\mu_{bc}^{IJ}(\kappa)\,\bigl(C\,c^{-(J)}\bigr)(\kappa)
\\
+\int_{-\infty}^{\infty}d\kappa\,b_0^{(I)}
\bigl[\mu^{IJ}_{bc}(\kappa)
-\mu_{bc}^{IJ}(0)\bigr]\,\mathscr{P}\frac{\sqrt{A_1(\kappa)}}{\kappa}\,
\bigl(C\,c^{-(J)}\bigr)(\kappa)
\biggr].
\label{Vgh2}
\end{multline}
From our experience with momentum Neumann matrices we know
that the zero modes can be added by a unitary transformation
\eqref{U}. We show that a similar thing happens
for ghosts: the term proportional to $b_0$ in
the exponent \eqref{Vgh2} can be obtained by a unitary
transformation acting on the first line of \eqref{Vgh2}.
The statement is
\begin{equation}
\langle V_3^{2,-1}|=N_{\textsf{gh}}\,{}_{123}\langle +|
\exp\left[
\int_{-\infty}^{\infty}d\kappa\,
b^{-(I)}(\kappa)\,\mu_{bc}^{IJ}(\kappa)\,\bigl(C\,c^{-(J)}\bigr)(\kappa)
\right]\,U_{\text{gh}}^{(1)}\otimes U_{\text{gh}}^{(2)}\otimes
U_{\text{gh}}^{(3)},
\label{Vgh3}
\end{equation}
\end{widetext}
where
the unitary operator $U_{\text{gh}}$ is given by
\begin{equation}
U_{\text{gh}}=\exp\left\{
-b_0\int_{-\infty}^{\infty}d\kappa\,
\mathscr{P}\frac{\sqrt{A_1(\kappa)}}{\kappa}\,
\bigl[c^{-}(\kappa)+c^{+}(\kappa)\bigr]
\right\}
\label{Ugh}
\end{equation}
The proof of this statement is very similar to the one
given in Section~\ref{sec:zeroNeumann}, one just has
to notice that \mbox{$\mu_{bc}^{IJ}(0)=+\delta^{IJ}$}.

Using \eqref{610b} and \eqref{xim}, one can rewrite
\eqref{Ugh} in the discrete
basis
\begin{equation}
U_{\text{gh}}=\exp\left\{
b_0\,\sum_{n=1}^{\infty}(-1)^n(c_{2n}+c_{-2n})
\right\}.
\label{Ugh2}
\end{equation}
This unitary operator and the
relation \eqref{Vgh3} have appeared before
in papers \cite{reduced,Erler}, though the details are different.
One should look on it
 as a \textit{unitary} redefinition of a string field $\Ac$. If
we redefine  $\Ac_{\text{new}}=U_{\text{gh}}\Ac_{\text{old}}$
then the interaction part of the cubic action simplifies, but
the kinetic term changes too
\begin{equation*}
Q_B\mapsto U_{\text{gh}}Q_B U^{-1}_{\text{gh}}.
\end{equation*}
In particular, in the Siegel gauge this new kinetic
operator becomes by \eqref{Ugh2}
\begin{equation}
\left.U_{\text{gh}}\, c_0 L^{\text{tot}}_0\, U^{-1}_{\text{gh}}
\right|_{b_0=0}
=\frac{1}{2i}\,\bigl[c(i)-c(-i)\bigr]L_0^{\text{tot}}.
\label{UcU}
\end{equation}
Notice that unlike $c_0$ the ghost piece of this
kinetic operator has a simple representation
in both $bc$ and bosonized formulations of the ghost CFT,
and resembles the conjecture of \cite{rastelli}.

\begin{figure}[!t]
\centering
\includegraphics[width=221pt]{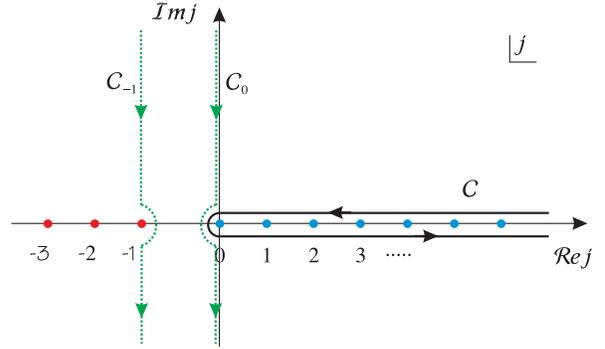}
\caption{The dots represent the poles
of the integrand in \eqref{int2}. Contour
$C$ encircles the positive real
axis counterclockwise. Then we deform it to
contour $C_0$ for the first term and to
contour $C_{-1}$ for the second term, which lie parallel to the
imaginary axis at $\Re j =0$ and $\Re j=-1$ correspondingly.}
\label{fig:2}
\end{figure}

\begin{widetext}
\subsection{$\beta\gamma$-superghosts}
The diagonalization of \eqref{superN} goes almost
in the same way as described in Section~\ref{sec:matter}.
So let us only sketch the derivation. Substitution
of the maps \eqref{h3}, expansion in a binomial series, and
turning the sum into a contour integral yields
\begin{multline}
M_{\beta\gamma}^{IJ}(z,\oz')=\cosh w\cosh\ow'\,\frac{1}{2i}
\oint_C\frac{dj}{\sin\pi j}\left\{
-\delta^{IJ}\,2i e^{\mp \frac{i\pi}{2}} \Bigl[e^{\pm i\pi}
e^{2(\ow'-w)}\Bigr]^{j+\frac12}
\right.
\\
\left.
-\frac{3}{8i}\,\Bigl[-e^{\frac{4}{3}(\ow'-w)+i(\varphi_J-\varphi_I)}\Bigr]^{j}
+\frac{3}{8i}\,\Bigl[-e^{\frac{4}{3}(\ow'-w)+i(\varphi_J-\varphi_I)}\Bigr]^{j+1}
\right\}
\label{int2}
\end{multline}
\end{widetext}
Now we want to deform the contour as shown on Figure~\ref{fig:2}
with $C_0$ for the first term and $C_{-1}$ for the second. But
before we do this we have to worry about falloff at infinity.
Using cyclic symmetry we choose $I=2$, then for $J=1,3$ we can interpret
\begin{equation*}
-e^{i(\varphi_J-\varphi_2)}=e^{\mp \frac{i\pi}{3}}.
\end{equation*}
After dividing by $\sin(\pi j)$ we
get the following asymptotic behavior of the integrand
\begin{equation*}
\sim e^{-2\pi|\Im j|/3},\quad\text{as}\quad \Im j\to\pm\infty.
\end{equation*}
Hence for $M_{\beta\gamma}^{2J}$ ($J\ne 2$) the
integrand vanishes at infinity. Instead for $J=I=2$ the term
with $\delta^{IJ}$ comes into play and cancels the integrand at infinity
(see details in Section~\ref{sec:3d}).
Finally we get
\begin{subequations}
\begin{align}
M_{\beta\gamma}^{2J}(z,\oz')&=\frac{1}{2}\,\cosh w\cosh\ow'\int_{-\infty}^{\infty}
d\kappa\,e^{i\kappa(w-\ow')}\,
\notag
\\
&\qquad\quad\times
\mathscr{P}\frac{e^{\mp\frac{\pi\kappa}{4}}}{\sinh\frac{3\pi\kappa}{4}}
\quad\text{for}\quad J=1,3;
\\
M_{\beta\gamma}^{22}(z,\oz')&=\frac{1}{2}\,\cosh w\cosh\ow'\int_{-\infty}^{\infty}
d\kappa\,e^{i\kappa(w-\ow')}\,
\notag
\\
&\qquad\quad\times
\left[
\mathscr{P}\frac{e^{\pm\frac{3\pi\kappa}{4}}}{\sinh\frac{3\pi\kappa}{4}}
\mp\frac{e^{\pm\frac{\pi\kappa}{2}}}{\cosh\frac{\pi\kappa}{2}}
\right].
\end{align}
\end{subequations}
The principal value comes from the sum of two integrals over contours
$C_{0}$ and $C_{-1}$ (see Figure~\ref{fig:2}), which now run on
opposite sides of the $\kappa=0$ pole. Comparison with
\eqref{kappas} shows that
we can interpret it as an expansion of the
$\beta\gamma$ Neumann matrix
in $s=\frac12$ $K_1$-eigenfunctions. Thus to obtain the eigenvalues
we have to use the $s=\frac12$ normalization \eqref{A1/2} of the eigenstates:
\begin{equation}
\begin{split}
\mu_{\beta\gamma}^{22}(\kappa) &= \pm\mathscr{P}\frac{\cosh \rx}{\sinh 3\rx},
\\
\mu_{\beta\gamma}^{21}(\kappa)&=e^{- \rx}\,\mathscr{P}\frac{\cosh 2\rx}{\sinh 3\rx},
\\
\mu_{\beta\gamma}^{23}(\kappa)&=e^{+\rx}\,\mathscr{P}\frac{\cosh 2\rx}{\sinh 3\rx},
\end{split}
\label{sgh}
\end{equation}
where $\rx\equiv\frac{\pi\kappa}{4}$.
The eigenvalue $\mu_{\beta\gamma}^{22}(\kappa)$ with the ``$+$'' sign
coincides with Arefeva et. al. \cite{arevefa} (eq. (7.11)).
The non-diagonal elements
$\mu_{\beta\gamma}^{I,I+1}$ and $\mu_{\beta\gamma}^{I+1,I}$
have to be switched in order to
agree with \cite{arevefa}. The origin of this switching
is related to the definition of the $3$-string vertex. Here
we use bra $3$-string vertex \eqref{Vsgh}, while
authors of \cite{arevefa} use ket $3$-string vertex (see
equation (7.9) therein). The relation between these two vertices
is precisely a switch $I\leftrightarrow I+1$ in \eqref{sgh}.

As in the case of the ghosts
\eqref{ghm}
one obtains the following relation between eigenvalues
of $3$-string matter fermion Neumann matrices \eqref{M1/2}
and picture $-1$ $\beta\gamma$-Neumann
matrices \eqref{sgh}
\begin{equation}
\mu^{IJ}_{\frac12}(\kappa)=\mu_{\beta\gamma}^{IJ}(\kappa\pm 2i).
\end{equation}
Notice that the skewsymmetry of the $s=\frac12$ Neumann
matrices \eqref{M1/2} appears automatically.

\section{Conclusion}
\label{sec:summary}
\setcounter{equation}{0}
    For nonzero scale dimension $s$, our results largely confirm
previous authors, though our proofs are much
shorter and clearer.
For $s>0$ we are dealing with unitary representations
of $SL(2,\Rh)$ or its covering group. The discrete basis
with $L_0$ diagonal is \eqref{basisn}. From it we constructed
a Cauchy kernel \eqref{id}, which projects
onto the entire Hilbert space.
A Watson-Sommerfeld transformation then expanded
it in the continuous $\kappa$-basis which diagonalizes $K_1=L_1+L_{-1}$.
The eigenfunctions are \eqref{kappas} with normalization
\eqref{As}. The transformation matrix between the bases is
\eqref{<mk>}.

Another integral kernel \eqref{Mmaps} generates the $N$-string
Neumann matrices \cite{peskin1}. The same Watson-Sommerfeld
transformation expands it in the
$K_1$ basis, where it is diagonal. The ratio to the
diagonalized Cauchy kernel then gives the Neumann
eigenvalues
\begin{subequations}
\begin{align}
\mu^{II}_{s,N}(\kappa)&=e^{\pi\kappa N/4}\frac{B_{s,N}(\kappa)}{A_s(\kappa)}
-e^{\pi\kappa/2},
\\
\mu^{IJ}_{s,N}(\kappa)&=e^{+\frac{\pi \kappa}{4}\,(N+2I-2J)}
\,\frac{B_{s,N}(\kappa)}{A_s(\kappa)}\quad(I<J),
\\
\mu^{IJ}_{s,N}(\kappa)&=(-1)^{2s}\,e^{-\frac{\pi \kappa}{4}\,(N-2I+2J)}
\,\frac{B_{s,N}(\kappa)}{A_s(\kappa)}\quad(I>J),
\end{align}
\end{subequations}
where $(-1)^{2s}$ reflects the symmetry or skewsymmetry
of the Neumann matrices ($\mu^{JI}_{s,N}(\kappa)=
(-1)^{2s}\,\mu_{s,N}^{IJ}(-\kappa)$), and
\begin{equation}
\frac{B_{s,N}(\kappa)}{A_s(\kappa)}=\left[\frac{2}{N}\right]^{2s-1}\,
\frac{\Gamma\Bigl(s+\frac{iN\kappa}{4}\Bigr)\Gamma\Bigl(s-\frac{iN\kappa}{4}\Bigr)}
{\Gamma\Bigl(s+\frac{i\kappa}{2}\Bigr)\Gamma\Bigl(s-\frac{i\kappa}{2}\Bigr)}.
\end{equation}
Note the simple formula valid for all $s$ and $N$.

Unitarity fails at $s=0$, where there is a vector with infinite norm.
In the second quantized world-sheet theory this corresponds
to a zero frequency oscillator,
and the divergence is eliminated by transforming it
to position and momentum (Section~\ref{sec:zeromodes}).
The Neumann eigenvalues are the same as for $s=1$, but the
eigenfunctions differ.
If $x,\,p,\,\alpha_n$ are the usual oscillators
from the $L_0$ basis \cite{GSW}, then the continuum
oscillators in the $K_1$ basis are
\begin{multline}
a^{\pm}(\kappa,p)
=\mp i \left[
\frac{\kappa}{2\sinh\frac{\pi\kappa}{2}}
\right]^{1/2}\left\{\mathscr{P}\frac{\hat{p}}{\kappa}
+\alpha_{\mp 1}+\frac{1}{2}\,\kappa\,\alpha_{\mp 2}
\right.
\\
\left.
+\frac{1}{6}(\kappa^2-2)\,\alpha_{\mp 3}+
\frac{1}{24}(\kappa^3-8\kappa)\alpha_{\mp 4}+\dots
\right\}.
\end{multline}
(We have suppressed Lorentz indices.)
For $p=0$ these reduce to the $s=1$ continuum oscillators.
The average position $x$ is also replaced by the midpoint
position
\begin{multline}
\xi=X_L(i)+X_L(-i)
\\
=x+i\sqrt{2\app}\,\sum_{n=1}^{\infty}
\frac{(-1)^n}{2n}\,\bigl[\alpha_{2n}-\alpha_{-2n}\bigr].
\end{multline}
Then $\xi,\,\hat{p},\,a^{\pm}(\kappa,p)$ form
a complete basis for the $s=0$ world-sheet field theory with
\begin{equation}
\begin{split}
&[\xi,\,\hat{p}]=i,\qquad
[a^{-}(\kappa,p),\,a^+(\kappa',p)]=\delta(\kappa -\kappa'),
\\
&
[\xi,\,a^{\pm}(\kappa,p)]=
[\hat{p},\,a^{\pm}(\kappa,p)]=0.
\end{split}
\end{equation}
$\xi$ and $\hat{p}$ arise from an additional nonnormalizable
state at $\kappa=0$. The expansion of $X_L(z)$ in
these new oscillators is \eqref{X_Lcont}. Similar
results for other world-sheet fields can be found in
\eqref{phipm} -- \eqref{psiz} and \eqref{610}.

Plane waves will now contain the midpoint position $\xi$
instead of $x$. The bosonized ghost insertions
at curvature points will therefore be very simple in
this basis.

The zero momentum and non-zero momentum oscillators
are related by the unitary transformation $U_p$
\begin{multline}
U_{p}=\exp\biggl\{
i\sqrt{2\app}\,\hat{p}\,\int_{-\infty}^{\infty}
d\kappa\,\lim_{s\to 0}\sqrt{A_s(\kappa)}\,
\\
\times\bigl[
a^{+}(\kappa,0)+a^{-}(\kappa,0)
\bigr]
\biggr\},
\end{multline}
where $\hat{p}$ is the momentum operator.
One can consider this alternatively as a unitary string field redefinition
$\Ac_{\text{new}}=U_p\Ac_{\text{old}}$.
There are two consequences.
Firstly, remember that the cubic interaction looks
nonlocal if it is written in the component
fields corresponding to $\Ac_{\text{old}}$ since it
involves exponentials of $p$.
This is cured by the field redefinition.
Secondly, the BRST charge changes to
\begin{equation*}
Q_B\mapsto U_p Q_B U_p^{-1},
\end{equation*}
which adds terms linear and quadratic in $p$.
Therefore the action in  the component fields
corresponding to $\Ac_{\text{new}}$ now looks local,
which may  help in constructing lump and rolling tachyon solutions.
A similar field redefinition \eqref{UcU} converts the ghost
zero mode $c_0$ into the conjectured kinetic term for the
nonperturbative vacuum \cite{rastelli}.

For the $bc$ and $\beta\gamma$ ghosts (Section~\ref{sec:ghost})
we took the easy way out by using a vacuum in which
their Neumann matrices can be related to those
of the matter fields. However the ghost eigenvalues
certainly depend on the
vacuum, and in other vacua are nonhermitian. This
question deserves further investigation,
as does BRST invariance in the $K_1$ basis.
In Appendix~\ref{app:L0} we suggest some expressions
for the Virasoro operator $L_0$ in the $K_1$ basis.

In usual field theory, $p$ space is appropriate
to weak coupling, $x$ space to strong coupling. Diagonalizing
the vertex may therefore allow a latticized strong
coupling approach to string field theory,
and make concrete the old idea of induced gravity.
Perhaps we really live in flat
$10D$ space-time, and what we see is just illusion.
This was one of our motivations for solving this preliminary
mathematical problem. Another motivation was
to study the relation of Witten's star
product to the Moyal product as described in \cite{Bars} and
\cite{moore}. This may help give a mathematical understanding
of the string algebra in the $K$-theory context \cite{Kt}.

\begin{acknowledgments}
We are grateful to H.~Liu, S.~Lukyanov and G.~Moore for useful discussions.
We would like to thank A.~Giryavets for
many valuable comments on the draft of this paper.
The work of D.M.B. was supported in part by RFBR grant 02-01-00695.
\end{acknowledgments}

\appendix
\renewcommand {\theequation}{\thesection.\arabic{equation}}

\section{$L_0$ in the $\kappa$-basis}
\label{app:L0}
\setcounter{equation}{0}
Here we calculate $L_0$ in the $\kappa$-basis.
By \eqref{Lnz} and \eqref{f->z} it takes
the following form in the $w$ coordinate
\begin{equation}
L_0=\frac{1}{2}\,\sinh(2w)\frac{d}{dw}+s.
\end{equation}
One can easily apply  this operator to the states
\eqref{kappas} which diagonalize $K_1$:
\begin{multline}
L_0|\kappa,s\rangle(w)
=\Biggl\{e^{2w}\left[\frac{s}{2}+\frac{i\kappa}{4}\right]
\\
+e^{-2w}\left[\frac{s}{2}-\frac{i\kappa}{4}\right]
\Biggr\}|\kappa,s\rangle(w).
\label{A2}
\end{multline}
From this and \eqref{kappas} one sees that $L_0$ is a
difference operator: it shifts $\kappa$ to $\kappa\mp 2i$ .
We can formally write down the kernel for this operator
\begin{multline}
\langle\kappa',s|L_0|\kappa,s\rangle=
\left[\frac{s}{2}+\frac{i\kappa}{4}\right]\,\delta(\kappa-2i-\kappa')
\\
+
\left[\frac{s}{2}-\frac{i\kappa}{4}\right]\,\delta(\kappa+2i-\kappa').
\end{multline}
Here $s$ is the scale dimension.
Notice the complex $\delta$-functions,
which occurred in previous papers \cite{dima2}
though details differ \footnote{
The complex $\delta$-function is a well known object
in mathematical physics \cite{vladimirov}, \cite{ruhl}. Its action on
holomorphic functions is defined by
$$
\int dz\,f(z)\delta(z-w)=\frac{1}{2\pi i}\int_{C_w} dz\,\frac{f(z)}{z-w},
$$
where contour $C_w$ encircles point $w$ in some way.
}.

The complex $\delta$-functions appeared because the
integrand does not fall off at infinity. This
suggests considering $L_0$ between vectors $|\kappa,s\rangle$ with
different $s$. Using standard Fourier transforms one finds
by \eqref{A2} and \eqref{kappas}
\begin{multline}
\langle\kappa',s+1|L_0|\kappa,s\rangle=
\left[\frac{A_s(\kappa)}{A_{s+1}(\kappa')}\right]^{1/2}\,
\\
\times
\left\{
-\mathscr{P}\,\frac{s(\kappa-\kappa')+\kappa}{2\sinh\frac{\pi(\kappa-\kappa')}{2}}
+2s\,\delta(\kappa-\kappa')\right\}.
\end{multline}
This formula gives a finite kernel for the operator $L_0$.

\section{Lemmas}
\label{app:lemma}
\setcounter{equation}{0}
Here we list some useful properties of the functions introduced
in Section~\ref{sec:eigen}.
By \eqref{As} and a standard Fourier transform
\begin{equation}
\int_{-\infty}^{\infty}d\kappa\,A_s(\kappa)e^{i\kappa w}
=\Gamma(2s)(\cosh w)^{-2s}.
\label{C1}
\end{equation}
By noticing that the kernels $\mathrm{Id}_s(z,\oz')$ in
equations \eqref{218} and \eqref{id} are the same and
expanding both equations in $z$ and $\oz'$ one concludes
\begin{equation}
\int_{-\infty}^{\infty}d\kappa\,A_s(\kappa)\,V_{m}^{(s)}(\kappa)
V_n^{(s)}(\kappa)=\bigl[N_m^{(s)}\bigr]^2\,\delta_{mn},
\label{C2}
\end{equation}
From this it follows that
the transition matrix \eqref{<mk>} is unitary.
Another way to obtain \eqref{C2} is to expand
\eqref{C1} as in \eqref{223}.
Differentiating \eqref{223}
with respect to $z$ and expanding
in $z$ one gets
\begin{equation}
\begin{split}
V_1^{(s)}(\kappa)&=\kappa\,V_{0}^{(s+1)}(\kappa),
\\
V_{m+1}^{(s)}(\kappa)&=\frac{1}{m+1}\,
\Bigl[\kappa\,V_{m}^{(s+1)}(\kappa)-2s\,V_{m-1}^{(s+1)}(\kappa)\Bigr].
\end{split}
\label{C3}
\end{equation}
Notice $s$ in the numerator. Because of it one gets the following
equation by \eqref{limits}
\begin{multline}
A_0(\kappa)V_{m+1}^{(0)}(\kappa)
=\frac{1}{m+1}
\biggl[A_1(\kappa)\,\mathscr{P}\frac{V_{m}^{(1)}(\kappa)}{\kappa}
\\
-V_{m-1}^{(1)}(0)\,\delta(\kappa)\biggr],
\label{C3'}
\end{multline}
where $\mathscr{P}$ means principal value.
By expanding \eqref{223} for $\kappa=0$ one obtains
\begin{equation}
V_{2m-1}^{(s)}(0)=0\quad\text{and}\quad
V_{2m}^{(s)}(0)=(-1)^m\,\frac{\Gamma(m+s)}{\Gamma(s)\Gamma(m+1)}.
\label{C3''}
\end{equation}
By dividing \eqref{223} by $s$
and taking  $s\to 0$
we obtain the following identity for  $\kappa=0$ and $m\geqslant 1$
\begin{equation}
\left.\frac{\pd V_{2m}^{(s)}(0)}{\pd s}\right|_{s=0}
=\frac{(-1)^m}{m}
\quad\text{and}\quad
\left.\frac{\pd V_{2m-1}^{(s)}(0)}{\pd s}\right|_{s=0}
=0.
\label{C4}
\end{equation}
By differentiating the recursion formula \eqref{req1}
\begin{equation}
\left.\frac{\pd V_{2m+1}^{(0)}}{\pd\kappa}\right|_{\kappa=0}=
\frac{(-1)^m}{2m+1},\quad\text{and}\quad
\left.\frac{\pd V_{2m}^{(0)}}{\pd\kappa}\right|_{\kappa=0}=0.
\label{C4.5}
\end{equation}
The recursion formula \eqref{req1} yields the following
representation for $V_{2n+1}^{(1)}$:
\begin{equation*}
\frac{V_{2n+1}^{(1)}(\kappa)}{\kappa}=\sum_{j=0}^{n}
(-1)^{n-j}\,\frac{V_{2j}^{(1)}(\kappa)}{2j+1},
\end{equation*}
and therefore by \eqref{C2}
\begin{equation}
\int_{-\infty}^{\infty}d\kappa\,A_1(\kappa)\,
\frac{V_{2n+1}^{(1)}(\kappa)}{\kappa}= (-1)^n.
\label{C5}
\end{equation}


\end{document}